%% file: main.tex
\documentclass[acmtog]{acmart}
\input{macros}

\copyrightyear{2026}
\acmYear{2026}
\setcopyright{cc}
\setcctype{by}
\acmConference[SA Conference Papers '26]{SIGGRAPH Asia 2026 Conference Papers}{December 01--04, 2026}{Kuala Lumpur, Malaysia}
\acmBooktitle{SIGGRAPH Asia 2026 Conference Papers (SA Conference Papers '26), December 01--04, 2026, Kuala Lumpur, Malaysia}
\acmDOI{10.1145/3829340.3842313}
\acmISBN{979-8-4007-2842-6/2026/12}

\begin{document}

\title{HoOPOE: A Holography-Optimized Perceptual ObjectivE function}

\ccsdesc[500]{Computing methodologies~Perception}
\ccsdesc[500]{Computing methodologies~Mixed / augmented reality}
\ccsdesc[300]{Hardware~Displays and imagers}

\keywords{Computer-Generated Holography, Perceptual Loss Functions, Image Quality Metrics, Augmented Reality}

\include{main-body}

\end{document}

%% file: macros.tex
\usepackage{graphicx}
\usepackage{xspace}
\usepackage{subcaption}
\usepackage{amsmath}

\ifx\figurename\undefined \def\figurename{Figure}\fi
\renewcommand{\figurename}{Fig.}

\newcommand{\Sect}[1]{Sec.~\ref{#1}}
\newcommand{\Fig}[1]{Fig.~\ref{#1}}

\newcommand{\Eqn}[1]{Eqn.~\ref{#1}}

\newcommand{\proj}{\textsc{HoOPOE}\xspace}

\newcommand{\no}[1]{}
\newcommand{\RNum}[1]{\uppercase\expandafter{\romannumeral #1\relax}}

%% file: main-body.tex
\input{sections/abstract}

\begin{teaserfigure}
    \centering
    \includegraphics[width=\linewidth]{newgraphs/Teaser.pdf}
    \caption{A comparison of a HoOPOE-trained 3D hologram against those trained under three other loss functions: Mean Squared Error (MSE), Mean Squared Error in LAB space (MSE LAB)~\cite{lab}, and SCCGH~\cite{sccgh}.
    The results shown here simulate temporal integration at 480 fps.
    Note that the subjective differences are more significant to an in-person observer using our physical setup, due to the differences in viewing conditions (e.g., spatial/temporal resolution, display luminance).
    A user study shows that our loss yields significantly better results than those of the other approaches. Copyright for the Blender scene (licensed under CC-BY-SA) used to generate this focal stack goes to  \citet{tada_blender_2023}.}
    \Description{Five side-by-side renderings of a bicycle Blender scene from \citet{tada_blender_2023}. From left to right: MSE, MSE LAB, SCCGH, \proj, and Reference renders.}
    \label{fig:teaser}
\end{teaserfigure}

\author{Ethan Chen}
\email{echen48@ur.rochester.edu}
\orcid{0009-0008-1250-769X}
\affiliation{
    \institution{University of Rochester}
    \city{Rochester}
    \state{New York}
    \country{USA}
}
\affiliation{
    \institution{Meta}
    \city{Sunnyvale}
    \state{California}
    \country{USA}
}

\author{Xi Peng}
\email{xipeng@cs.unc.edu}
\orcid{0009-0004-5759-7917}
\affiliation{
    \institution{University of North Carolina at Chapel Hill}
    \city{Chapel Hill}
    \state{North Carolina}
    \country{USA}
}

\author{Alexandre Chapiro}
\email{achapiro@meta.com}
\orcid{0000-0002-7367-0131}
\affiliation{
    \institution{Meta}
    \city{Sunnyvale}
    \state{California}
    \country{USA}
}

\author{Dario Seyb}
\email{dario.seyb@gmail.com}
\orcid{0000-0002-0234-8401}
\affiliation{
    \institution{Meta}
    \city{Redmond}
    \state{Washington}
    \country{USA}
}

\author{Michael Grundmann}
\email{mgrundmann@meta.com}
\orcid{0009-0007-2434-961X}
\affiliation{
    \institution{Meta}
    \city{Menlo Park}
    \state{California}
    \country{USA}
}

\author{Carl Marshall}
\email{csmarshall@meta.com}
\orcid{0009-0001-7288-5341}
\affiliation{
    \institution{Meta}
    \city{Redmond}
    \state{Washington}
    \country{USA}
}

\author{Sushant Kondguli}
\email{sushant.kondguli@gmail.com}
\orcid{0000-0002-7295-4626}
\affiliation{
    \institution{Meta}
    \city{Sunnyvale}
    \state{California}
    \country{USA}
}

\author{Praneeth Chakravarthula}
\email{cpk@cs.unc.edu}
\orcid{0000-0002-3092-7435}
\affiliation{
    \institution{University of North Carolina at Chapel Hill}
    \city{Chapel Hill}
    \state{North Carolina}
    \country{USA}
}

\author{Yuhao Zhu}
\email{yzhu@rochester.edu}
\orcid{0000-0002-2802-0578}
\affiliation{
    \institution{University of Rochester}
    \city{Rochester}
    \state{New York}
    \country{USA}
}

\maketitle

\input{sections/introduction}

\input{sections/related}

\input{sections/architecture}

\input{sections/benchmarking}
\input{sections/discussion}

\input{sections/conc}

\begin{acks}
The work is partially supported by NSF
Awards \#2225860, \#2126642, and \#2044963 as well as a Meta research grant.
\end{acks}

\input{sections/last2pages}

\clearpage

\bibliographystyle{ACM-Reference-Format}
\bibliography{refs.bib}

%% file: sections/abstract.tex
\begin{abstract}
Holographic devices must reconstruct an entire light field given a finite number of spatial light modulator (SLM) pixels, and must do so using scalar-wave diffraction from coherent illumination alone.
As a result, computer-generated holography (CGH) often has speckling artifacts.
We propose \proj (Holography-Optimized Perceptual ObjectivE function), a CGH-specific, perceptually-aware loss function that hides speckles from human vision.
\proj is effective for CGH optimizations because it accounts for the human visual system's sensitivity to spatio-temporal frequencies and color, and its parameters are tuned to maximize its performance as a loss function for CGH. 
We use \proj to generate holograms for mesh-derived focal stacks.
We then conduct a user study to benchmark \proj-generated holograms against existing loss functions.
\proj yields the most visually pleasing results; users prefer \proj 68\% of the time over the next best objective function on average.
\end{abstract}

%% file: sections/introduction.tex
\section{Introduction}

Holographic displays have many advantages over traditional emissive displays for virtual and augmented reality.
Conventional displays present flat images at a fixed focal length, relying only on binocular disparity to generate depth cues.
In contrast, holographic displays can theoretically replicate the entire outgoing light field by using a spatial light modulator (SLM) to manipulate the wavefront of light emitted from coherent (laser) sources.
This allows holographic displays to provide the correct depth cues to drive accommodation, a key factor in immersive 3D experience.
Additionally, holographic displays have naturally wide color gamuts, can support larger eyeboxes, and can eliminate bulky waveguides and projective lenses.

However, the quality of images rendered on holographic displays is harmed by speckle noise, which arises from the constructive and destructive interference of coherent light emitted by the  holographic display.
Speckle noise obscures fine detail, reduces contrast, and introduces distracting visual artifacts that undermine the advantages that holographic displays would otherwise offer.
Unlike many visual artifacts that can be solved through calibration or post-rendering correction, speckle is intrinsic to holographic displays.

The classic approach to speckle reduction is to increase the degrees of freedom in the system.
This is achieved through either temporal multiplexing, which requires operating the SLM at a high frame rate, or through a high-resolution SLM.
The increased degree of freedom allows us to more easily synthesize a proper spatio-temporal SLM pattern to match a given wavefront.
Neither approach is desirable, however, as they significantly increase the computational cost and power consumption of the system.

We take the reverse perspective and ask: without increasing the system complexity, how can we most effectively hide speckles from human vision --- guided by a perceptual loss function?

Prior perceptual metrics such as CVVDP~\cite{cvvdp}, MILO~\cite{cogalan2025milo}, LPIPS~\cite{zhang2018perceptual}, and DISTS \cite{ding2020image}, while effective as quality predictors, may not be effective as objective functions for CGH algorithms.
These metrics are often trained and calibrated on image quality datasets~\cite{ding2021comparison},
where natural images are corrupted with a specific set of artifacts (e.g., blur, brightness changes, color saturation, spatial non-uniformity).
When using these metrics as loss functions for CGH, the optimizer will discover stimuli that were unseen during metric calibration.
``Adversarial examples'' can easily ``hack'' the perceptual metrics:
trained stimuli have low loss but are visually objectionable (see \Fig{fig:adversarial} and Supplementary Material~S11 for examples).

Existing perceptual metrics, understandably, perform suboptimally on these types of stimuli, because these stimuli are only produced during CGH training (when the metrics are used as a loss function) and are not commonly present in existing image quality datasets used for training metrics.
There are no large-scale subjective datasets available that would allow for a thorough recalibration.

In this work, we introduce \proj (Holography-Optimized Perceptual ObjectivE), a loss function designed specifically for holographic image optimization.
Our approach is guided by a key requirement: effective perceptual loss functions must not only correlate with human perception, but must also remain robust against adversarial stimuli encountered during optimization.
\proj accounts for the contrast sensitivity of human vision to stimuli of varying spatiotemporal frequencies and color, and is specifically optimized for stimuli that one might encounter during hologram synthesis.

Using a gradient-descent CGH algorithm~\cite{zhang20173d, liu2020optimization, peng2020neural} as a case study, we show that \proj generates more visually pleasing holograms compared to existing loss functions.
The loss function, by being temporally aware, also informs the design of time multiplexing: the biggest perceptual gains occur when increasing the frame rate from 120 frames per second to 240 fps.
The increase in perceptual quality for the jump between 240 fps and 480 fps was not statistically significant.

We make the following contributions:

\begin{itemize}
    \item We introduce \proj, a novel perceptual loss function for CGH that accounts for various aspects of human vision, while remaining robust during optimization.
    \item We evaluate the performance of state-of-the-art perceptual quality metrics, and demonstrate limitations when used as holographic optimization losses.
    \item We benchmark \proj against existing loss functions using a pairwise comparison psychophysical test. We show that \proj generates holograms that are significantly more visually pleasing. 
\end{itemize}

%% file: sections/related.tex
\section{Related Work}

\subsection{Perceptual Loss Functions for CGH}

The shift from alternating-projection methods to gradient-based optimization \cite{chakravarthula2019wirtinger, peng2020neural} reframed CGH as a question of \textit{what to optimize} rather than \textit{how to optimize}. 
As it is impossible to remove all speckles from scalar wave diffraction alone using coherent illumination \cite{iwai1996speckle, goodman2007speckle}, perceptual models can play a critical role in improving visual quality by hiding speckles where they are least visible.

\citet{yang2022perceptually} conducted a comprehensive benchmark of image quality metrics as holographic objective functions. 
The authors evaluated 12 metrics using a two-alternative forced choice test, and found that MS-SSIM~\cite{msssim} outperformed all alternatives.
However, their study was restricted to monochromatic image content.
The metrics they benchmarked did not account for chromatic or temporal aspects of perception.

For color holography, Simultaneous Color Computer Generated Holography (SCCGH) \cite{sccgh} proposed a novel objective function that calculates loss in a cone opponent color space. 
SCCGH applies a low-pass filter to the image to eliminate theoretically invisible signals of high spatial frequency.
While this approach models both color perception and spatial contrast sensitivity, it does not account for temporal contrast sensitivity.

\begin{figure*}
    \begin{subfigure}{0.24\textwidth}
        \includegraphics[trim={420 0 420 0},clip,width=\linewidth]{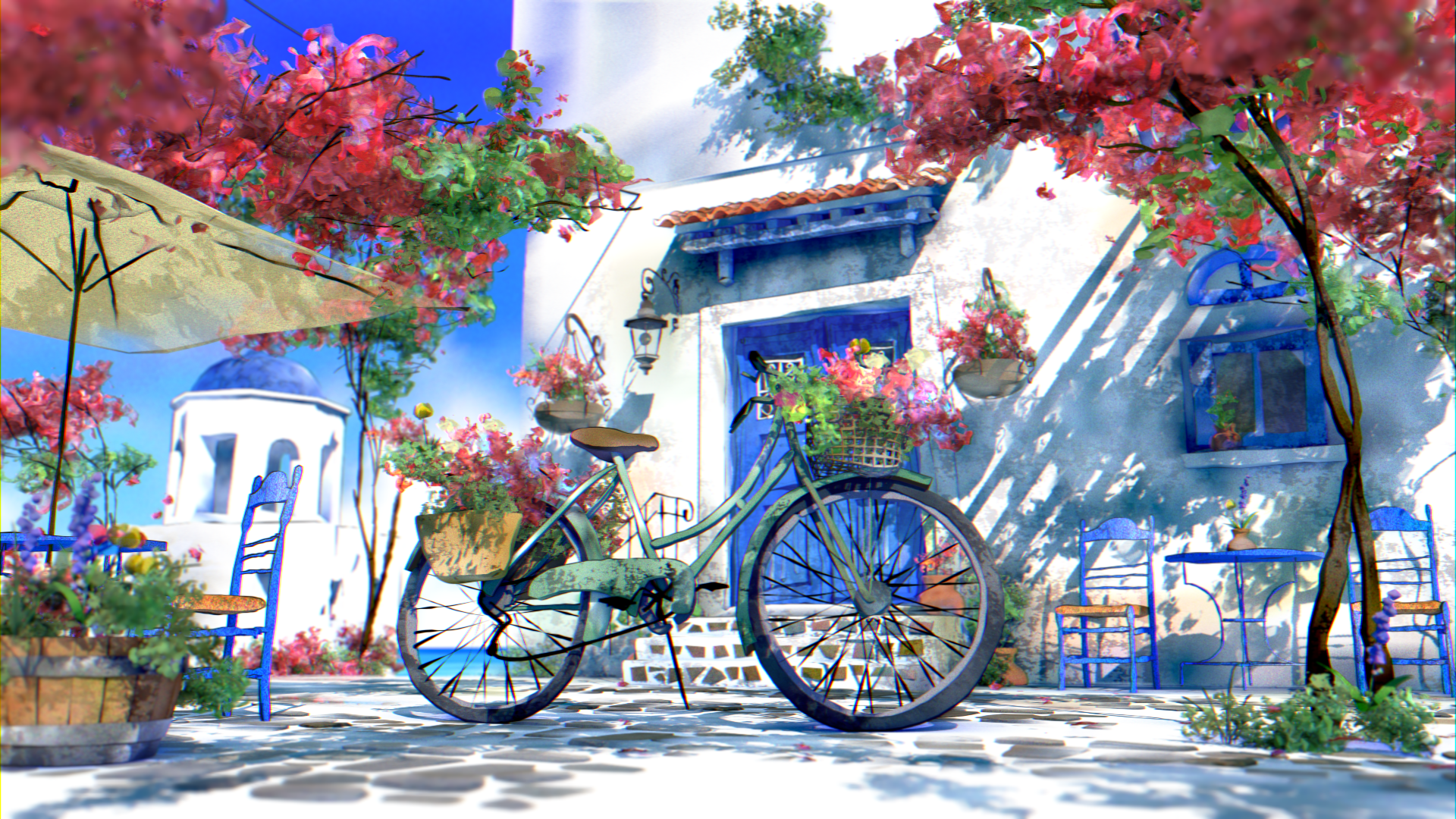}
        \caption{Reference image. JOD: 10}
    \end{subfigure}
    \begin{subfigure}{0.24\textwidth}
        \includegraphics[trim={420 0 420 0},clip,width=\linewidth]{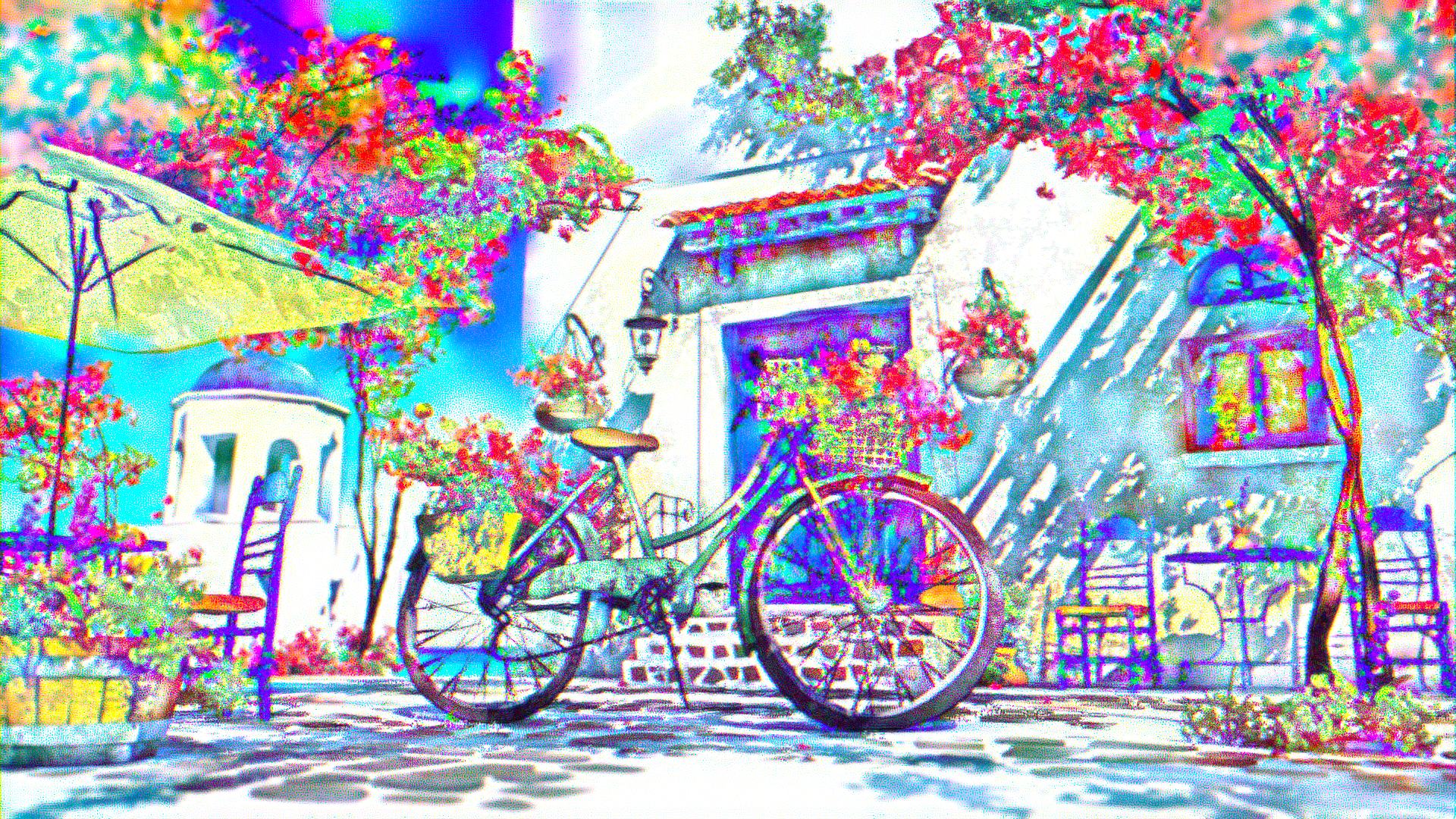}
        \caption{Predicted JOD: 8.51}
        \label{fig:adversarial:cvvdp}
    \end{subfigure}
    \begin{subfigure}{0.24\textwidth}
        \includegraphics[trim={420 0 420 0},clip,width=\linewidth]{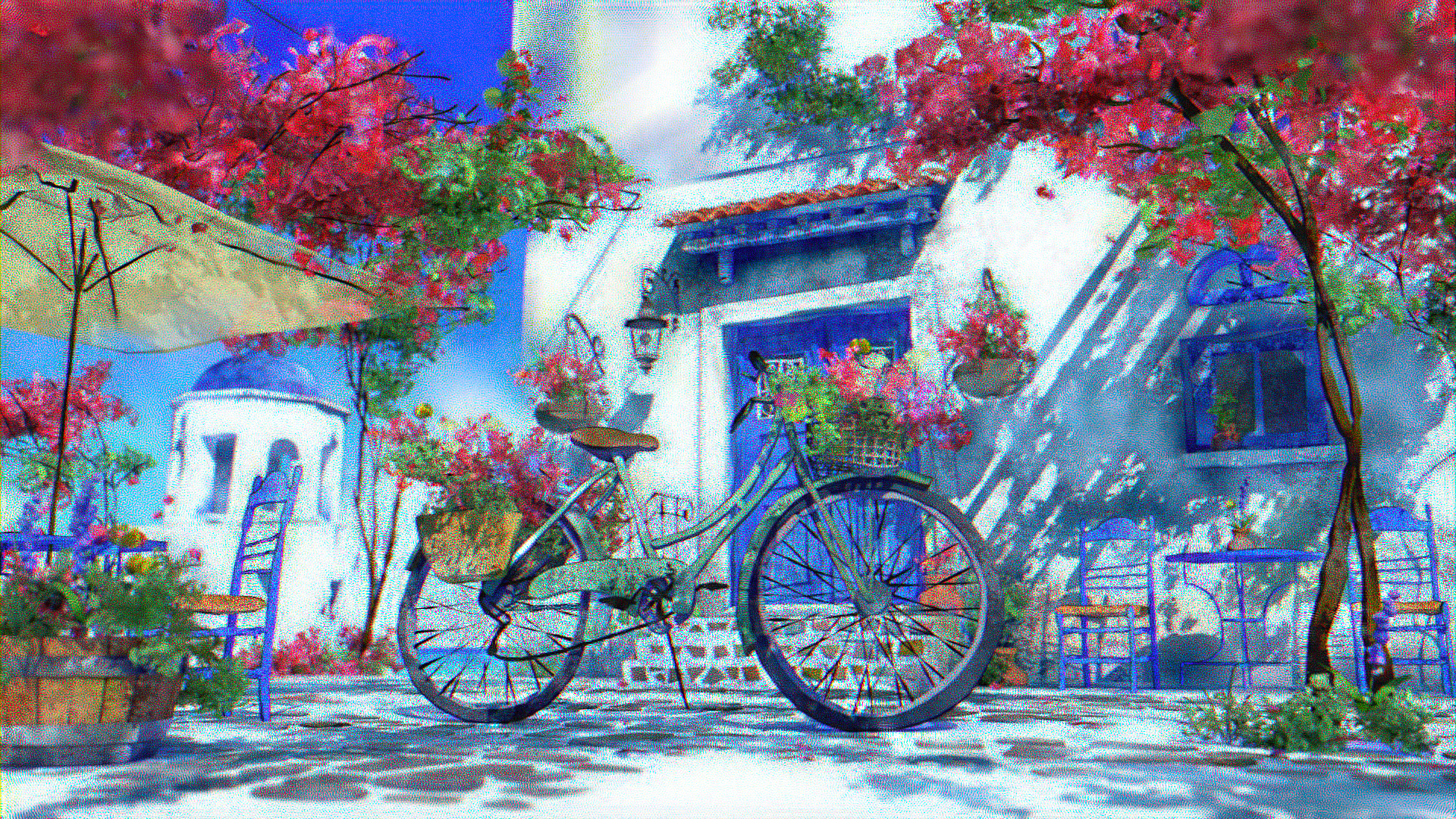}
        \caption{Predicted JOD: 7.03}
        \label{fig:adversarial:hoopoe}
    \end{subfigure}
    \vspace{-5pt}
    \caption{
    CVVDP scores \Fig{fig:adversarial:cvvdp} (trained by CVVDP) over 1 JOD higher than \Fig{fig:adversarial:hoopoe} (trained by \proj), even though \Fig{fig:adversarial:hoopoe} is visually preferable.
    CVVDP's standard full HD display model was used to score both images. See Supplementary Material~S11 for more examples.
    Copyright for the original 3D model (licensed CC-BY-SA) goes to \citet{tada_blender_2023}.
    }
    \Description{Three side by side images of different renderings of a bicycle. The leftmost image is the original (unmodified) rendering of the bicycle. The middle image is a CVVDP-trained CGH reconstruction of the scene. The rightmost image is a \proj-trained reconstruction of the scene. The CVVDP version has a higher (better) JOD score, but looks visibly discolored compared to the \proj-trained alternative.}
    \label{fig:adversarial}
\end{figure*}

Others explored gaze-contingent techniques to optimize holographic rendering. 
\citet{chakravarthula2021gaze} exploits differences in peripheral and foveal contrast sensitivity to suppress speckle more aggressively in foveal vision.
\citet{walton2022metameric} extends \citet{freeman2011metamers} and \citet{walton2021beyond} to holography by formulating a loss function that accounts for ventral metamerism.
\citet{chakravarthula2022pupil} further improves holographic rendering by accounting for the pupil size and position of the viewer.
These methods must be computed in real-time, as they rely on eye tracking data.
While they do account for variations in spatial contrast sensitivity, these solutions are not temporally aware.

Not every CGH algorithm can accommodate perceptual loss functions, however. 
Closed-form solutions, like \citet{maimone2017holographic}, \citet{choi_gaussian_2025}, and \citet{padmanaban2019holographic}, sidestep optimization altogether.
Because no objective is minimized, the choice of loss function does not arise in these pipelines.

\subsection{Temporal Multiplexing in Holography}

Temporal multiplexing has emerged as a practical approach for speckle reduction.
Uncorrelated speckle noise displayed at a sufficiently high frame rate becomes perceptually invisible, as the human visual system integrates across time.

\citet{takaki2011speckle} and \citet{Mori:14} decomposed an image into a set of point groups, then displayed them sequentially to recreate the original image.
\citet{choi2022time} further advanced temporal multiplexing by demonstrating that jointly optimizing hologram frames across time significantly increased the image reconstruction quality.
The authors calculate loss by time-averaging the holographic image to simulate the temporal summation of the human eye.
\citet{sha2024multi} and \citet{lee2022high} also tabulate loss against the time-averaged holographic image.

However, existing works on temporal multiplexing in holography share a common limitation. 
They do not account for the human visual system's variable sensitivity across temporal frequencies.
Prior work makes the implicit assumption that all noise that is above a specific temporal frequency threshold is equally imperceptible with an instantaneous cutoff. 
In reality, contrast sensitivity functions show that sensitivity declines gradually with increasing temporal frequency \cite{ashraf2024castlecsf, Robson:66}.
This suggests that weighting errors across temporal frequency bands could yield perceptual benefits beyond what current methods achieve.

\subsection{Perceptual Metrics vs. Objective Functions}

\begin{figure*}
    \centering
    \includegraphics[width=\textwidth]{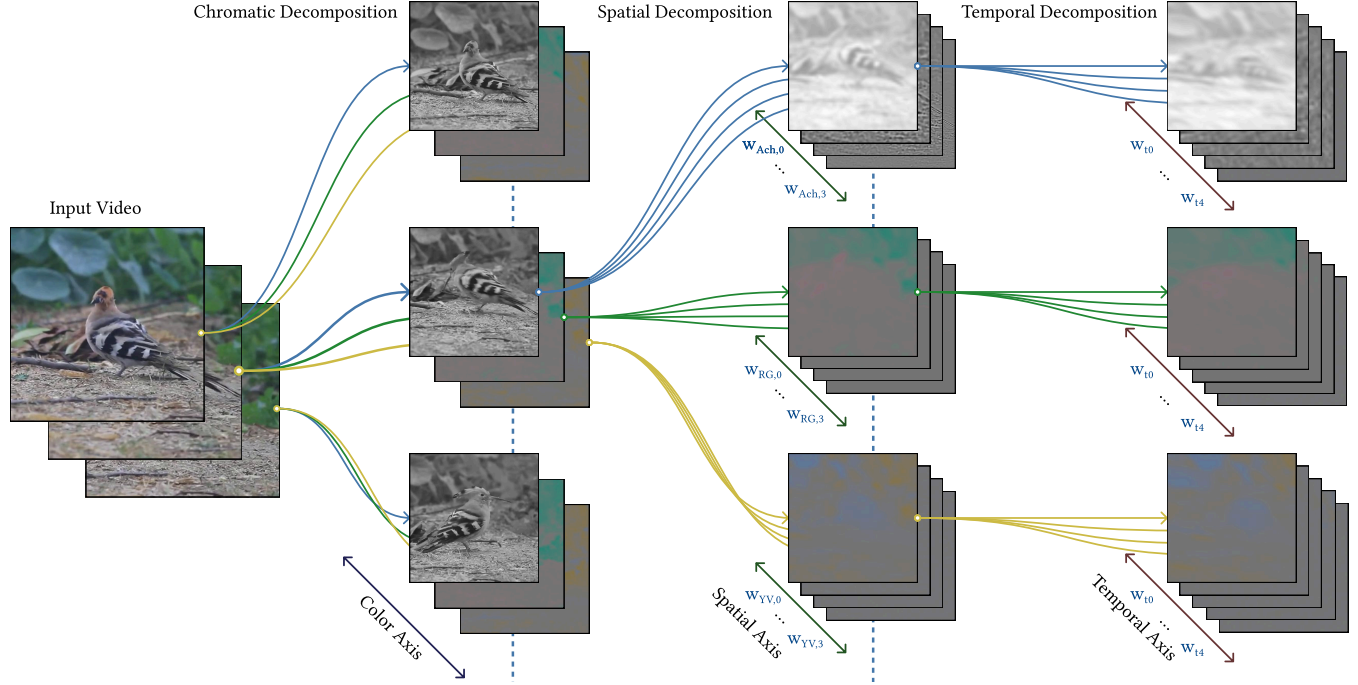}
    \caption{Visualization of HoOPOE's architecture.
    We decompose each video frame into three chromatic channels.
    Each channel is then decomposed into a set of $N$ spatial frequency bins through a Laplacian pyramid.
    Each combination of color channel $c$ and spatial-frequency $s$ is associated with a trainable weight $w_{c, s}$.
    For each $c$ and $s$ bin, we further decompose the content into a set of $M$ temporal-frequency bins through a Laplacian pyramid.
    Each temporal-frequency bin $t$ is associated with a trainable weight $w_t$.
    The final loss is a weighted MSE across the bins between the reference and test videos.
    To avoid visual clutter, only one spatial band per channel is temporally decomposed in this graphic.
    Copyright for the original video (licensed CC-BY) goes to \citet{ram_eurasian_2018}.
    }
    \Description{A visualization of the \proj pipeline. An input video is first fed into \proj. This input video first undergoes chromatic decomposition, where each frame is decomposed into three distinct channels, one for each of the L*, a*, and b* axes of CIELAB space \cite{lab}. Then each channel is further decomposed into a Laplacian pyramid, where each level of the pyramid represents a different spatial frequency band. Finally, the spatially and chromatically decomposed frame undergoes temporal decomposition via Laplacian pyramid as well, further separating the spatial chromatic channel by temporal frequency.}
    \label{fig:hoopoe:architecture}
\end{figure*}

Image quality metrics evolved from simple pixel-wise comparisons to sophisticated models of human perception.
Early structural similarity methods like SSIM \cite{ssim} and MS-SSIM \cite{msssim} compare local luminance, contrast, and structure between images.
NLPD \cite{laparra2017perceptually} computes perceptual distance through modeling contrast and gain control via Laplacian pyramid decomposition.
More recent learned metrics like LPIPS \cite{zhang2018perceptual} are trained to be strong predictors of human perceptual judgments.
Visible Difference Predictors (VDPs) \cite{cvvdp, mantiuk2021fovvideovdp} are perceptual quality metrics that account for contrast sensitivity across different spatial and temporal frequencies, color channels, and eccentricities.

Many of these metrics perform well on image quality benchmark datasets like TID2013 \cite{ponomarenko2015image}. 
However, a good predictor is not necessarily a good loss function, and optimizing for perceptual metrics may not yield visually pleasing images~\cite{mustafa2022training}.
\citet{ding2021comparison} conducted a study demonstrating this gap, showing that
many metrics with good evaluation performance were outperformed by MSE when used as loss functions for image generation tasks.
Similarly, we observe that when used as an objective function, CVVDP generates ``adversarial stimuli'' that achieve high predicted quality scores, but results may still appear visibly degraded (\Fig{fig:adversarial} and Supplementary Material~S11).

Complex perceptual metrics can be highly non-convex, hurting convergence~\cite[Sec. 6.2]{cvvdp}.
Interestingly, for speckle suppression, we find that the non-convexity of existing metrics is \textit{not} the primary reason they perform poorly as loss functions.
For example, in Supplementary Material~S11 we show that although the optimization landscape of CVVDP contains many basins and the optimizer can be trapped in different ones depending on initialization, these basins tend to achieve similarly low loss values.
We observe the same behavior for other metrics, consistent with prior observations in learning theory research~\cite{du2018gradient, allen2019convergence}.
Instead, the ineffectiveness of existing metrics such as CVVDP as a loss function for CGH is caused by ``adversarial stimuli'' whose quality is systematically misjudged by the metric.

%% file: sections/architecture.tex
\section{HoOPOE}
\label{sec:hoopoe}

\proj is a loss function grounded in human vision and tailored for computer-generated holography.
Our loss function takes as input a reference video (ground truth) and a test video (generated by a CGH algorithm), and outputs a single loss value representing the perceptual difference between the two.
We provide an overview first, followed by elaborations on the design details.

\subsection{Overview}
\label{sec:hoopoe:ov}

Our loss function is designed with three goals in mind.
First, it must be perceptually grounded.
To that end, \proj converts both the reference and test videos into an opponent color space approximating the opponent process of color vision.
Each color channel (achromatic, yellow-blue, and red-green) is then decomposed into a set of bins consisting of varying spatial and temporal frequency bands.
Operating in the spatio-temporal-chromatic space rather than the RGB space allows us to, through learned weighting later, account for the human contrast sensitivity to varying spatial and temporal frequencies across color channels.

Second, the loss must facilitate the smooth, gradient-based optimization that CGH algorithms usually rely on.
To that end, we compute the loss as a simple weighted mean squared error (MSE) across the spatio-temporal-chromatic bins between the reference and test videos.
Specifically, each spatio-temporal-chromatic bin is assigned a trainable weight;
we calculate the difference between the reference and test videos on each weighted bin, and pool the per-bin difference together to form the loss using MSE.
A visual representation of our loss function's pipeline can be found in \Fig{fig:hoopoe:architecture}.

Formally, the HoOPOE loss function is defined as
\begin{equation}
    l_{\text{HoOPOE}} = \sum_{c, s, t} \frac{w_{c, s} \cdot w_t}{\| P_{s,t,c} \|} \sum_{p \in P_{s, t,c}} (L_{\text{ref}}(p,c,s,t) - L_{\text{test}}(p,c,s,t))^2.
    \label{eq:hoopoe_main}
\end{equation}

\noindent where $L(p, c, s, t)$ denotes the magnitude at pixel $p$ in color channel $c$, spatial Laplacian pyramid band $s$, and temporal Laplacian pyramid band $t$.
$P_{s, t, c}$ is the set of all pixels across all spatio-temporal bands and chromatic channels.
$w_{c, s}$ and $w_t$ are trainable weights representing the perceptual importance of each spatio-chromatic bin and temporal bin, respectively.

The loss formulation produces a quadratic (convex) objective function in the prediction error, with well-conditioned gradients that support efficient optimization. 
At the same time, the trainable parameters $w_{c, s}$ and $w_t$ permit perceptually motivated and CGH-specific weighting across bins, which leads us to our third point.

Finally, the loss must be robust against adversarial stimuli encountered during optimization. 
These stimuli minimize the loss while being perceptually objectionable.
The key idea is that the spatio-temporal-chromatic weights in the loss function should be tuned with the CGH algorithm in the loop.
In this way, the loss function can correctly score the perceptual quality of content with CGH-specific artifacts.
This CGH-in-the-loop loss function tuning amounts to a challenging bi-level optimization problem.
Solving that is the goal of this work.

\subsection{Color Space Conversion}
\label{sec:hoopoe:cs}

We first convert the input video data into the CIELAB color space~\cite{lab}, used as the basis for our loss function.
The CIELAB color space was derived with the goal of perceptual uniformity.
Calculating color difference in a perceptual color space is critical for a loss function to produce perceptually accurate images.

Compared to more recent color spaces designed for perceptual uniformity, such as CAM16 UCS~\cite{cam16}, ICtCp~\cite{ictcp}, and OKLab~\cite{oklab}, CIELAB uses a simpler transformation from and to display-referred color spaces, such as sRGB.
In practice, we found that this reduces the  complexity of the optimization landscape and leads to better convergence.
Alternative physiologically relevant color spaces such as the color-opponent DKL space were not employed. 
As they do not aim to be perceptually uniform, the gradients in this space are less useful for guiding optimization.

\subsection{Spatial and Chromatic Weighting}
\label{sec:hoopoe:spatial}

We then decompose each color channel into a set of $N$ spatial-frequency bands through a Laplacian pyramid.
Each pyramid level acts as a band-pass filter, representing a particular spatial frequency band.
We associate each of the $3 \times N$ pyramid levels with a weight $w_{c, s}$, which determines the relative importance of each spatial frequency band $s$ and color channel $c$ in the perceptual loss.
We fit the weights specifically for computer-generated holograms according to their perceptual quality using the following procedure.

\paragraph{Optimization Space.}
We consider $N=4$ to reduce the optimization space. As such, the set of weights $\pmb{w} = \{ w_{c, s}\}$ is composed of 12 scalars, corresponding to four frequency bands and three color channels such that each $w_{c, s} \in (0, 1]$.
We also tune two learning parameters: learning rate and step decay. 

\paragraph{Optimization Problem.}
At a high level, we perform a bi-level optimization.
The goal is to optimize a set of weights $\pmb{w}$ of \proj (outer problem) such that the perceptual quality is maximized when a holographic image is trained under \proj with $\pmb{w}$ (inner problem).
This bi-level optimization problem is formulated as follows:
\begin{subequations}
\label{eq:hoopoe}
\begin{align}
    \pmb{w}^* &= \underset{\pmb{w}}{\mathrm{argmax}} \left[ l_{\text{scoring}} \left( r, \text{CGH}(\theta_{\pmb{w}}^*) \right) \right] \label{eq:hoopoe:outer}, \\
    \text{where} \quad \theta^*_{\pmb{w}} &= \underset{\theta}{\mathrm{argmin}} \left[ l_{\text{HoOPOE}} \left( r, \text{CGH}(\theta); \pmb{w} \right) \right].
    \label{eq:hoopoe:inner}
\end{align}
\end{subequations}

In the outer optimization, $\pmb{w}$ is the set of all 12 $w_{c, s}$ parameters, and $l_{\text{scoring}}$ is a scoring function (see below) that scores the perceptual quality, with respect to the reference image $r$, of a test image generated by the CGH algorithm using the SLM pattern $\theta_{\pmb{w}}^*$.
In the inner optimization problem, we are optimizing the SLM pattern $\theta$ to minimize the HoOPOE loss parameterized by the parameters $\pmb{w}$. 

\paragraph{Reference Image Selection.}
We select a random set of images from Segment Anything 1B \cite{kirillov2023segment} as hologram optimization targets.
Five images in total were selected and cropped to 1080p.

\paragraph{Optimization Strategy.}
While the inner optimization problem in \Eqn{eq:hoopoe:inner} can be solved using first-order, gradient-based methods, the outer optimization problem in \Eqn{eq:hoopoe:outer} only has the final value of the inner optimization problem, and cannot compute gradients of the outer loss back to $\pmb{w}$.

Therefore, we use a zeroth-order optimization algorithm CMA-ES \cite{cmaes}, implemented in the popular framework Optuna \cite{optuna}, to optimize our spatial-color weights $\pmb{w}$.
The low dimensionality of $\pmb{w}$ also makes the search space friendly to zeroth-order techniques.
It is theoretically possible to solve \Eqn{eq:hoopoe:outer} and \Eqn{eq:hoopoe:inner} in tandem using hypergradient descent \cite{hyperparameter}, which we leave for future work.

Specifically, we repeatedly alternate between adjusting the loss function weights via Optuna and performing CGH optimization via gradient descent in a loop.
For each set of weights $\pmb{w}$ proposed by Optuna, we form a loss function, using which we run a complete gradient descent hologram optimization pass using \proj as a loss function.
This produces a holographic image.
We ran 1300 tuning trials in total.

\paragraph{Perceptual Scoring.}
We have two options to form $l_{scoring}$ in \Eqn{eq:hoopoe:outer}, which scores the quality of CGH content: 1) human-in-the-loop, where humans score CGH stimuli interactively as they are being generated, or 2) leveraging existing perceptual quality metrics such as CVVDP.
The former is intractable; we estimate 200 hours per participant for human-in-the-loop optimization.

Although CVVDP was not suitable as a direct loss function to optimize SLM pixels, motivating the development of \proj, it performs well as a perceptual scoring function for optimizing $w_{c, s}$. 
When CVVDP is used as a loss function,
the optimizer will necessarily encounter ``adversarial stimuli'' that maximize CVVDP's predicted quality while remaining subjectively objectionable (see \Fig{fig:adversarial} and Supplementary Material~S11 for examples).
As a consequence, we found the resulting gradients unreliable in guiding the optimization.
In contrast, we employ CVVDP to generate the 12 weights in the \proj loss.
By reducing the degrees of freedom of the optimization problem (from millions of pixels to only 12 parameters), we reduce the possibility of adversarial configurations appearing in the parameter search space.

\paragraph{Interpretation.}
Recent perceptually-based methods use models of human contrast sensitivity function (CSF) to generate weights for local contrast bands~\cite{cvvdp, mantiuk2021fovvideovdp, tariq2023perceptually, tursun2019luminance},
which has been shown to produce good results when comparing test and reference stimuli at similar luminance levels.
We show that our weighting method can be interpreted as a \textit{learned} approach toward contrast weighting.
We discuss in Supplementary Material~S2 why existing CSF-based weighting does not directly apply.

Using the model of~\citet{peli1990contrast}, the local Weber contrast of a pixel $p$ can be expressed by dividing the pixel value in a given Laplacian pyramid level by the corresponding pixel value in the Gaussian pyramid two levels below it in resolution:
\begin{align}
    \label{eq:c1}
    C(p, s) = \frac{L(p, s)}{G(p, s+2)},
\end{align}
\noindent where $C(p, s)$ is the local contrast of pixel $p$ at level $s$, $G(p, s+2)$ is the pixel value at the Gaussian pyramid level $s+2$, which represents the local adapting/mean luminance, and $L(p, s)$ is the pixel value at the Laplacian pyramid level $s$, which represents the local variation from the local mean.
$C$, $G$, and $L$ are each defined per color channel $c$, but from this point on we drop the channel index $c$ in the notation for simplicity's sake.

Given that $L(p, s) = G(p, s) - G(p, s+1)$ and $L(p, s+1) = G(p, s+1) - G(p, s+2)$, we know $L(p, s) + L(p, s+1) = G(p, s) - G(p, s+2)$.
Therefore, \Eqn{eq:c1} can be re-written as:
\begin{align}
    \label{eq:c3}
    C(p, s) = \frac{L(p, s)}{G(p, s) - (L(p, s) + L(p, s+1))}.
\end{align}
Now that each Laplacian level is associated with a weight, we can express a form of weighted contrast as:
\begin{subequations}
\label{eq:c4}
\begin{align}
    C_w(p, s) &= \frac{w_{c, s}L(p, s)}{G(p, s) - (w_{c, s}L(p, s) + w_{c, s+1}L(p, s+1))} \\
    &= \frac{G(p, s) - (L(p, s) + L(p, s+1))}{G(p, s) - (w_{c, s}L(p, s) + w_{c, s+1}L(p, s+1))} \cdot C(p, s).
\end{align}
\end{subequations}
Therefore, weighting the Laplacian pyramid levels has the effect of weighting local contrast of pixels, where the contrast weights are dependent on both the local image statistics (e.g., $G(p, s)$ and $L(p, s)$) as well as the learned weights assigned to each Laplacian level (e.g., $w_{c, s}$).
Compared to conventional contrast weighting (Eqn.~5b in Supplementary Material~S2), our weighting provides two degrees of freedom to weight the contrast of each pixel\footnote{These two degrees of freedom are shared by all the pixels at the same pyramid layer.} learned in accordance with perceptual quality.

\subsection{Temporal Weighting}
\label{sec:hoopoe:temporal}

We can also extend \proj to account for the temporal dimension of human vision.
For each combination of spatial-frequency band and chromatic channel, we build a temporal Laplacian pyramid with $M$ layers, each of which is associated with a weight $w_t$.
We use a value of $M=3$ to obtain a fine-grained decomposition of temporal frequency, which is needed to explore the effect of temporal multiplexing commonly used in CGH.
The set of weights $\pmb{w}_t = \{w_t\}$ determines the relative importance of errors in different temporal frequency bands of the reference and test videos.

The temporal weighting is considered in isolation from the spatio-chromatic weighting to reduce the parameter space.
This is supported by the fact that spatial and temporal filtering in retinal processing are approximately separable, or are modeled as such (for chromatic channels)~\cite{ashraf2024castlecsf}.
As a result, there are $3 \times N \times M$ bins, but only $3 \times N + M$ trainable weights.

We determine $w_t$ through a human-in-the-loop procedure.
Specifically, we first use a Sobol sampler to draw 50 different temporal weight configurations, each of which is used to train the CGH algorithm and produce a corresponding set of stimuli.
We then apply a single-elimination tournament approach to select the weighting that yields the most preferred video output.
Unlike spatial weighting, human-in-the-loop selection is viable here, as $\pmb{w}_t$ is composed of a set of 3 weights, as opposed to the 12 weights of $\pmb{w}_{c, s}$.
The parameter search space grows exponentially as a function of parameter count, so $\pmb{w}_t$'s search space is 9 orders of magnitude smaller than $\pmb{w}_{c, s}$.

%% file: sections/benchmarking.tex
\section{Validation}
\label{sec:validation}

We conduct an IRB-approved user study to compare \proj against prior loss functions for CGH.

\subsection{Stimuli}
\label{sec:validation:generation}

We use \proj to generate 5 CGH stimuli for subjective validation.
We generate a focal stack of images derived from Blender's demo scene files.
For each demo scene, we generated 5 images from a Blender camera with 5 different focal distances.
These images have no overlap with the images used to fit the \proj weights, and can be found in Supplementary Material~S5.

\subsubsection{CGH Stimuli Generation.}

We use a forward model based on the Angular Spectrum Method (ASM) optimized through gradient descent~\cite{chakravarthula2019wirtinger, chakravarthula2022differentiable, peng2020neural}.
Specifically, we use the ASM propagation model from \citet{peng2020neural}'s repository, modified to incorporate a Fourier aperture of 0.5.
More details can be found in Supplementary Material~S20.
We use SGD rather than neural network-based generation to isolate the effect of loss function choice on final image quality.
Neural architectures like \citet{shi2021towards}, \citet{peng2020neural}, and \citet{liu2025propagation} could potentially introduce inductive biases like network capacity that would obscure the effect of loss function on visual quality.

We make the following assumptions:

\begin{itemize}
    \item We optimize a focal stack for 5 propagation distances between 0.5 cm and 2.5 cm, uniformly spaced
    \item Phase-only SLM with a resolution of $1920 \times 1080$
    \item Pitch size of 8 $\mu$m
    \item RGB laser wavelengths of 638 nm, 520 nm, and 450 nm
\end{itemize}

For each RGB frame of the ground truth image, we generate three independent SLM patterns, one per laser wavelength. 
We do not perform any ROI cropping.

\subsubsection{Baselines.}
In addition to \proj, we benchmark three other loss functions in our user study.
All loss functions, including \proj, had their learning rate and step decay parameters tuned using Optuna~\cite{optuna} to ensure that any difference in quality stems from the loss function and not from hyperparameter selection.
Latency and memory overhead for training under each loss function are reported in Supplementary Material~S18.

\paragraph{Simultaneous Color Computer Generated Holography}
While not the central focus of their paper, \citet{sccgh} (SCCGH hereafter) proposes a perceptual loss function.
The key idea of SCCGH's loss function is to filter out theoretically invisible high spatial frequency features before using MSE loss on the filtered stimuli.
The original implementation in the paper uses empirical models of human vision.
We implement an enhanced version using more psychophysically accurate models of human vision.
See Supplementary Material~S1 for more details.

\paragraph{Mean Squared Error.}
We also benchmark the conventional MSE metric, which is a simple yet common metric used in prior CGH literature~\cite{yang2022perceptually, peng2020neural, choi2022time}.
In particular, we find that forming the MSE in the holographic display's color space (as opposed to other color spaces) performs the best.

\paragraph{Ablation.}
We perform an ablation study on HoOPOE by testing each addition in the HoOPOE pipeline in our validation study.
We test the following losses as part of our ablation study:
\begin{itemize}
    \item MSE tabulated in CIELAB~\cite{lab}. This represents HoOPOE in the absence of any frequency weighting (labeled MSE LAB).
    \item \proj with spatial-chromatic weightings only, without temporal frequency weighting (labeled Spatial \proj).
\end{itemize}

\paragraph{Temporal multiplexing.}

All tested conditions except \proj optimize the eight frames of the stimulus independently, relying on stochastic decorrelation for speckle reduction.
\proj is the exception, as \proj's mathematical formulation considers temporal frequency when tabulating loss (\Sect{sec:hoopoe:temporal}), and thus jointly optimizes all eight frames together.
We do not adopt the time-averaged optimization technique of \citet{choi2022time} for the baseline loss functions as it is incompatible with the conventional display we perform validation on.
When loss is tabulated on a time-averaged image, individual frames are free to leave our OLED's gamut, so long as the average image is close to the reference stimulus.
These out-of-gamut pixels, when gamut-clipped and displayed sequentially, result in visible discoloration. 
This discoloration is an artifact of our display, not of the hologram itself. 
For reference, $31\%$ of pixels are out-of-gamut for jointly optimized 8-frame MSE, as opposed to $1\%$ of pixels when the frames are independently optimized.

\paragraph{Other Losses.}
We chose these loss functions as they empirically performed the best compared to a range of other alternatives that we have experimented with, including SSIM~\cite{ssim}, MS-SSIM~\cite{msssim}, MILO~\cite{cogalan2025milo}, ColorVideoVDP \cite{cvvdp}, MSE in CAM16 UCS~\cite{cam16}, and MSE in OKLAB~\cite{oklab}.
We include examples of stimuli generated with these loss functions in Supplementary Material~S3.

\paragraph{Non-iterative CGH}

We chose to exclude non-iterative holography techniques (like those described in \citet{maimone2017holographic}, \citet{padmanaban2019holographic}), as these frameworks are not compatible with perceptual loss functions.
Subsequent work~\cite{peng2020neural, chakravarthula2022hogel, chakravarthula2019wirtinger} has reported higher reconstruction quality for iterative, gradient-based methods.

\subsection{Psychophysical Task}
\label{sec:validation:task}

\begin{figure}
    \includegraphics[width=\linewidth]{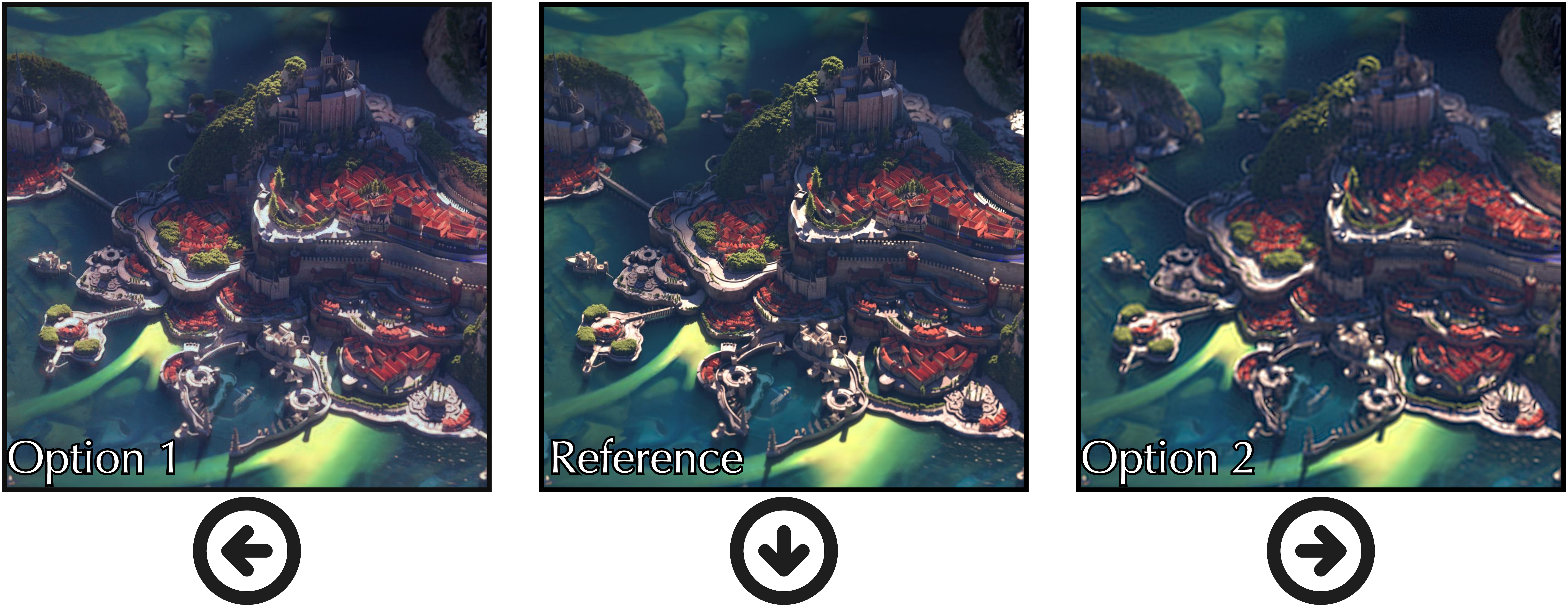}
    \caption{An example of a 2IFC task. Participants use arrow keys to switch between three stimuli -- two CGH renders and one ground truth stimulus. After viewing all three stimuli, the participant is allowed to select which one of the two options they perceive to be a more faithful recreation of the ground truth. This methodology is similar to \citet{chen2024pea}.
    Images are cropped to $1080 \times 1080$ here, but are full HD in the actual study.
    Copyright for the original 3D model (licensed CC-BY-SA) goes to \citet{krynski_blender_2022}.
    }
    \Description{Three overhead images of a castle, presented side by side. The leftmost rendition of the castle has the text "Option 1" overlaid upon it. The center-most image is labeled "Reference." The rightmost rendering is labeled "Option 2." Underneath each of these renderings is an arrow key, with the left arrow key corresponding to Option 1, the down key corresponding to Reference, and the right arrow key corresponding to Option 2.}
    \vspace{-15pt}
    \label{fig:validation:2afc}
\end{figure}

Each loss function was evaluated at four frame rates: 60, 120, 240, and 480 fps.
This yields a total of 21 conditions per image (5 loss functions $\times$ 4 frame rates, plus ground truth).
Each stimulus consisted of an 8-frame holographic video. 
Frames were looped continuously.

\paragraph{Procedure.}
\label{sec:psych:procedure}
We use a 2-Interval Forced Choice (2IFC) task for the psychophysics;
a visualization of the task is in \Fig{fig:validation:2afc}.
Participants are presented with three stimuli: a central reference stimulus and two comparison stimuli.
The stimuli are rendered on the center $1920 \times 1080$ pixels of the screen.
Participants switch between which stimuli they are viewing using keyboard inputs.
After the participant views all three stimuli, they are prompted to input which one of the two comparison stimuli they perceive to be closer to the reference stimulus.
This psychophysical task is similar to \citet{chen2024pea}.
The two comparison stimuli are generated by randomly selecting two of the 21 possible conditions, which are then randomly assigned to either side.
Despite optimizing the hologram for five planes, only the simulated image at 1.5 cm is shown to the user. 
We perform this simplification to reduce the complexity of the 2IFC task.
The entire set of generated stimuli can be found in Supplementary Material~S7.

\paragraph{Adaptive Sampling.}
Since we are benchmarking 5 scenes over 21 conditions, 6,300 pairwise comparisons are required per participant, which, if done naively, would take roughly 9 hours per participant.
We utilize the ASAP~\cite{asap} method to sample pairwise comparisons to maximize information gain.
This cuts down the time needed to obtain a good fit to under an hour per participant.

\paragraph{Participants.}
We recruit 12 participants for our validation study (ages 20--30, 1:1 male female ratio).
All have normal or corrected-to-normal vision.
None had been involved in the research project.

\paragraph{Hardware setup}

Stimuli were presented on a Sony INZONE M10S OLED display, which supports refresh rates up to 480 Hz. The display was configured to render SDR sRGB images, with a white point luminance of 275 nits. 
Participants were comfortably seated 0.7 meters from the display.
Head position was stabilized using a chin rest to ensure consistent viewing distance.

\subsection{Results}
\label{sec:results}

\begin{figure*}[ht]
    \includegraphics[width=.625\textwidth]{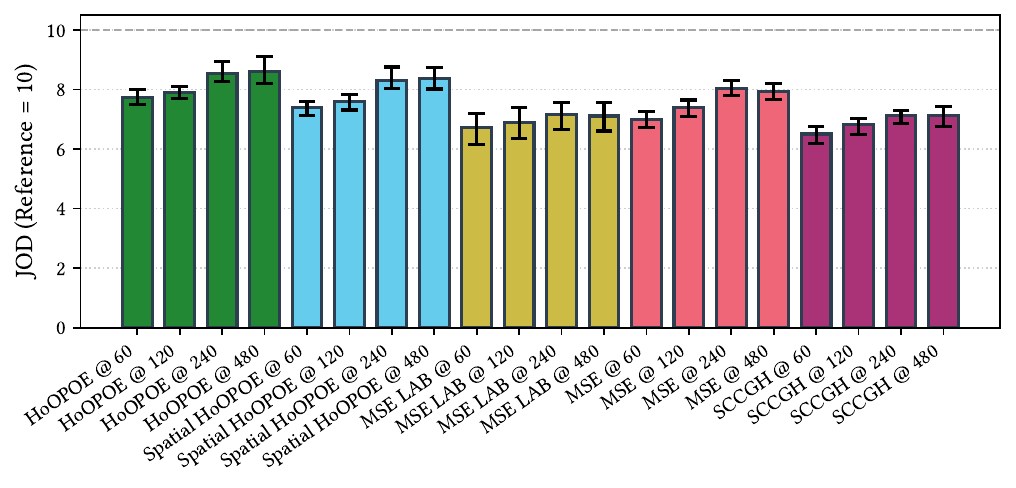}
    \vspace{-10pt}
    \caption{JOD values tabulated with pwcmp~\cite{pwcmp} over 4,362 comparisons. The Y axis represents JOD values. A difference of 1 JOD corresponds to a 75\% chance of preferring one stimulus over another. Higher JOD values are better. JOD values by stimuli are in Supplementary Material~S8.}
    \Description{JOD values and uncertainty bars for the benchmarked loss functions. From left to right: \proj, Spatial \proj, MSE LAB, MSE, SCCGH. Each benchmarked loss function is further split into separate FPS columns: 60 fps, 120 fps, 240 fps, and 480 fps.}
    \label{fig:validation:results}
\end{figure*}

A total of 4,362 pairwise comparisons were performed across 12 participants.
Pairwise comparison results were converted into Just Objectionable Difference (JOD) scores via the pwcmp library~\cite{pwcmp}.
1 unit of JOD corresponds to a 75\% chance of a participant preferring one stimulus over another.
We used 1,000 bootstrap samples to generate the fit.
One participant was classified as an outlier by pwcmp and excluded from the data analysis (dist\_L of $2.0$, which exceeds the customary threshold of 1.5).

\Fig{fig:validation:results} compares the JOD value across all methods (reference has a JOD of 10) across all stimuli; Supplementary Material~S8 shows the JOD results for each individual stimulus.
\Fig{fig:examples} visually compares different methods using one stimulus;
Supplementary Material~S7 shows more results of other stimuli.
Sample SLM patterns for the benchmarked loss functions are in Supplementary Material~S16. 

Our method significantly outperforms other benchmarks at their respective frame rates ($p < 0.05$, with some $p < 0.001$). 
Supplementary Material~S9 shows the hypothesis test results.
Spatial HoOPOE outperforms MSE LAB, which demonstrates the effectiveness of spatial and chromatic weighting. 
HoOPOE outperforms Spatial HoOPOE, which demonstrates the value of temporal weighting.

While more performance can be gained from increasing the frame rate of the holographic display, there are diminishing returns. 
The jump from 120 fps to 240 fps represents the greatest improvement in visual quality.
Users prefer the 240 fps rendering over the 120 fps rendering 74\% of the time.
However, users are only 50.7\% likely to prefer the 480 fps rendering over the 240 fps rendering.
Since display power consumption is a function of refresh rate, our results point to opportunities to make a calculated power-vs-quality trade-off.

To confirm that \proj's advantage is not dependent on temporal multiplexing, we conduct a user study (N=18) with single-frame stimuli. 
\proj continues to significantly outperform baseline losses under these conditions (Supplementary Material~S15).

%% file: sections/discussion.tex
\section{Discussion}
\label{sec:disc}

\subsection{Generalizability of \proj}
\label{sec:discuss:generalize}
The fundamental methodology of \proj generalizes to different hardware setups 
and viewing conditions.
Weights are tuned once for each setup, which takes about 6 hours.
Given that hardware and viewing distances are fixed for an XR device, this one-time overhead is negligible.
We demonstrate that \proj generalizes to different hardware setups and data distributions, even without tuning.

First we show that \proj generalizes to a different ASM model~\cite{matsushima2009band}.
\proj still outperforms all baselines (Supplementary Material~S10).
Additionally, we re-tune HoOPOE's spatial-chromatic weights for two different SLM setups and propagation distances (4.5 $\mu$m with 0.25 cm to 1.25 cm distances, 12.5 $\mu$m with 2 cm to 10 cm distances).
HoOPOE still outperforms all other loss functions in both scenarios, even without temporal weight tuning (Supplementary Material~S13).
We also generate holographic images under the aforementioned two SLM hardware setups without re-tuning \proj.
A user study (N=18) demonstrates that users still prefer \proj-generated stimuli over those of other losses. Results can be found in Supplementary Material~S14.
HoOPOE also generalizes beyond the validation dataset.
HoOPOE empirically outperforms other losses in optical-see-through UI content (Supplementary Material~S19).

\proj can generalize across different hardware configurations and viewing conditions because the tuning procedure represents images in device-independent units (e.g., retinal frequency) using CVVDP.
Specifically, CVVDP serves as the scoring function used to optimize the \proj weights (\Sect{sec:hoopoe:spatial});
it takes as input a display profile that specifies the characteristics of the display, and uses this information to convert the \proj-generated stimuli into display-independent measures of frequency, luminance, and color.

In our current tuning procedure (\Sect{sec:hoopoe:temporal}), we utilize humans-in-the-loop to set \proj's temporal weights. 
In a pilot study, we found that multiple judges converged to the same set of temporal weights, given the same set of Sobol-sampled temporal weight configurations.
Despite being tuned by a small number of participants on SA-1B data~\cite{kirillov2023segment}, the temporal weights generalize to the validation study performed on a different dataset (Blender renders) and on a large number of (different) participants (N=30 in \Sect{sec:validation} and Supplementary Material~S14). This suggests participant- and content-wise generalization for the temporal weighting scheme.

\subsection{Validation on a Conventional Display}
\label{sec:discussion:conventional}

\begin{figure*}
    \includegraphics[width=0.746\linewidth]{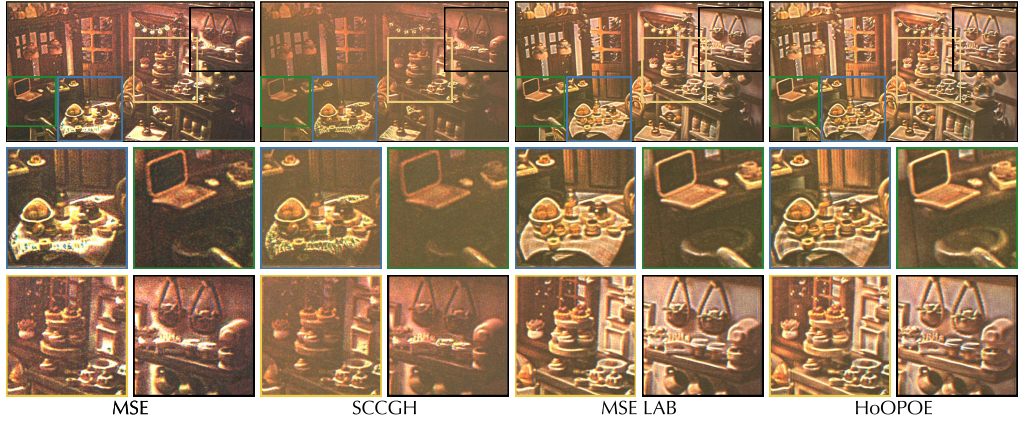}
    \vspace{-10pt}
    \caption{Camera captures for MSE, SCCGH, MSE LAB, and Spatial HoOPOE. MSE LAB is more color-accurate than MSE. HoOPOE preserves contrast, especially in dark regions. Only 1 frame (3 sequential RGB frames) was optimized and displayed. SLM patterns rendered in Fig.~S72. More details are discussed in Supplementary Material~S12. Copyright for the original 3D model (licensed under CC-BY-SA) goes to \citet{morena_blender_2023}.}
    \Description{Four hardware capture images for a CGH rendering of a kitchen. From left to right: MSE, SCCGH, MSE LAB, and \proj. Each hardware capture has four zoomed in views, which represent the same section of the image across all four loss functions.}
    \label{fig:camcap}
\end{figure*}

We chose to perform validation on a conventional display rather than a holographic prototype.
The goal of \proj is to mask artifacts from the propagation model itself.
As is well-known in the literature \cite{peng2020neural, lut}, an actual CGH display would introduce confounding display-specific artifacts.
Validating on a conventional display isolates the contribution of the loss function itself, and is a well-established approach in literature~\cite{georgiou2023visual, yang2022perceptually, chakravarthula2021gaze}.
HoOPOE is also complementary with prior techniques that address display-specific artifacts (e.g. CITL training~\cite{peng2020neural}):
HoOPOE can serve as the loss function for such pipelines, where the learned forward model can absorb display-specific distortions.

Simulation-based evaluation on a calibrated conventional display additionally makes our results display-agnostic and reproducible.
That said, we performed real hardware studies on a CGH display using the HOLOEYE PLUTO-2.1 SLM.
We empirically find that \proj still outperforms all baselines judging by camera captures of the display output.
An example is shown in \Fig{fig:camcap}, and more examples and details are discussed in Supplementary Material~S12.

\subsection{Tone Enhancement in \proj}

Because \proj weights loss across different Laplacian pyramid layers, \proj-reconstructed images may over- or under-emphasize high frequency edges if the weights are improperly tuned.
This could make images appear blurred or tone enhanced.
However, our 2IFC study allows direct comparisons of image regions against the reference.
Artifacts like tonal enhancement and blurring are immediately visible under 2IFC. Users still prefer our method over alternatives, suggesting that the artifacts brought by \proj do not negatively impact user experience.
In Supplementary Material~S21, we also analyze the randomness of \proj's SLM patterns.

\subsection{HoOPOE as a Perceptual Metric}
\label{sec:discuss:perceptualmetric}
HoOPOE was designed as a loss function, not as a perceptual metric. 
As observed by \citet{mustafa2022training}, good domain-specific loss functions are not necessarily good visual quality predictors. 
HoOPOE adopts the same task-specificity principle for CGH:
rather than reusing existing metrics~\cite{cvvdp, cogalan2025milo}, we instead fit HoOPOE weights against CGH-specific artifacts.

That said, HoOPOE still performs well as a perceptual metric when used on a video quality dataset, XR-DAVID \cite{xrdavid}, \textit{without} any retuning.
We outperform SCCGH, MSE LAB, MS-SSIM, SSIM, and LPIPS~\cite{zhang2018perceptual} in predicting stimuli quality.
A table of Spearman's rank correlation coefficients can be found in Supplementary Material~S6.

\subsection{Modeling of Other Visual Phenomena}
\label{sec:discuss:modeling_other}

HoOPOE aims to maximize performance as a loss function by only modeling aspects of human vision that are beneficial for optimization.
We experimented with visual masking, local contrast calculation, and contrast sensitivity in \proj.
These additions degraded results, and were excluded from the user study.
A comparison of these HoOPOE variants is in Supplementary Material~S4.

\subsection{\proj-Applicable Optimization Targets}
\label{sec:applicability}

\proj compares reference and test frame data in image space.
It is agnostic to how image data is generated. 
In \Sect{sec:validation:generation} the focal stack was derived from 3D mesh data.
However, \proj is applicable to any CGH pipeline that optimizes an image-space objective, including 3D~\cite{choi2022time, lee2022high}, pupil-aware~\cite{chakravarthula2022pupil}, and light field based~\cite{schiffers2023stochastic} approaches.
\proj is independent of the scene representation and propagation model that produce the images.

In our user studies we benchmark \proj's performance using static reference images, but \proj is applicable to video content as well.
The temporal decomposition in \Eqn{eq:hoopoe_main} is performed on both the test and reference stimuli.
We demonstrate that \proj is applicable to video reference data generated from a different underlying 3D representation format in Supplementary Material~S17.

%% file: sections/conc.tex
\section{Conclusion}

We propose \proj, a loss function for computer-generated holography.
The main contribution is a methodology that is robust against ``adversarial stimuli'' encountered during CGH optimization that have low loss under existing perceptual metrics but are subjectively objectionable.
\proj addresses this by 1) parameterizing the loss function across the spatio-temporal-chromatic dimensions that are perceptually relevant, and 2) tuning \proj parameters with a CGH algorithm in the loop.
We show that CGH content generated using \proj is visually preferable to content generated using existing perceptual metrics.

%% file: sections/last2pages.tex
\begin{figure*}
    \includegraphics[width=0.9\textwidth]{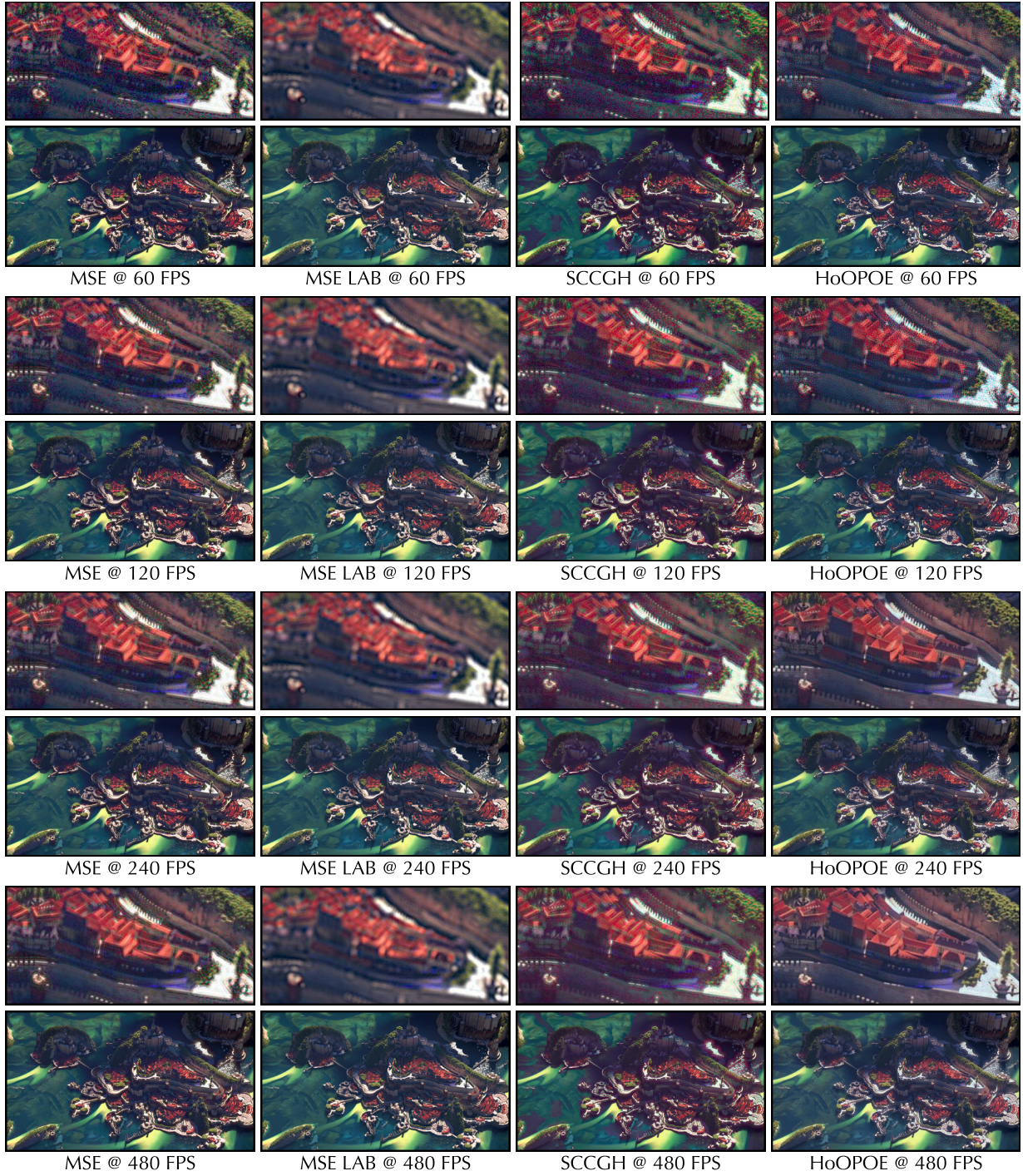}
    \caption{Images generated by HoOPOE, MSE, MSE LAB, and SCCGH at various frame rates for stimulus 1. The ground truth of this stimulus is Fig.~S5. Other examples can be found in Supplementary Material~S7. Note that this visualization of time multiplexing via time averaging is only an approximation of real-life viewing. Temporal artifacts like flicker are invisible in this render. Copyright for the original 3D model (licensed under CC-BY-SA) goes to \citet{krynski_blender_2022}.}
    \Description{A 4x4 grid of renderings of an overhead scene of a castle surrounded by water. The X axis of the grid represents loss function. The four displayed loss functions are MSE, MSE LAB, SCCGH, and \proj. The Y axis of the grid represents framerate. 60 fps, 120 fps, 240 fps, and 480 fps renders are displayed. Alongside each render is a zoomed in view of a small section of the castle rendering depicting some houses with red roofs.}
    \label{fig:examples}
\end{figure*}

%% file: main.bbl

\begin{thebibliography}{63}


\ifx \showCODEN    \undefined \def \showCODEN     #1{\unskip}     \fi
\ifx \showISBNx    \undefined \def \showISBNx     #1{\unskip}     \fi
\ifx \showISBNxiii \undefined \def \showISBNxiii  #1{\unskip}     \fi
\ifx \showISSN     \undefined \def \showISSN      #1{\unskip}     \fi
\ifx \showLCCN     \undefined \def \showLCCN      #1{\unskip}     \fi
\ifx \shownote     \undefined \def \shownote      #1{#1}          \fi
\ifx \showarticletitle \undefined \def \showarticletitle #1{#1}   \fi
\ifx \showURL      \undefined \def \showURL       {\relax}        \fi
\providecommand\bibfield[2]{#2}
\providecommand\bibinfo[2]{#2}
\providecommand\natexlab[1]{#1}
\providecommand\showeprint[2][]{arXiv:#2}

\bibitem[Akiba et~al\mbox{.}(2019)]%
        {optuna}
\bibfield{author}{\bibinfo{person}{Takuya Akiba}, \bibinfo{person}{Shotaro
  Sano}, \bibinfo{person}{Toshihiko Yanase}, \bibinfo{person}{Takeru Ohta},
  {and} \bibinfo{person}{Masanori Koyama}.} \bibinfo{year}{2019}\natexlab{}.
\newblock \showarticletitle{Optuna: A Next-generation Hyperparameter
  Optimization Framework}. In \bibinfo{booktitle}{\emph{Proceedings of the 25th
  ACM SIGKDD International Conference on Knowledge Discovery \& Data Mining}}
  (Anchorage, AK, USA) \emph{(\bibinfo{series}{KDD '19})}.
  \bibinfo{publisher}{Association for Computing Machinery},
  \bibinfo{address}{New York, NY, USA}, \bibinfo{pages}{2623–2631}.
\newblock
\showISBNx{9781450362016}
\href{https://doi.org/10.1145/3292500.3330701}{doi:\nolinkurl{10.1145/3292500.3330701}}


\bibitem[Allen-Zhu et~al\mbox{.}(2019)]%
        {allen2019convergence}
\bibfield{author}{\bibinfo{person}{Zeyuan Allen-Zhu}, \bibinfo{person}{Yuanzhi
  Li}, {and} \bibinfo{person}{Zhao Song}.} \bibinfo{year}{2019}\natexlab{}.
\newblock \showarticletitle{A Convergence Theory for Deep Learning via
  Over-Parameterization}. In \bibinfo{booktitle}{\emph{Proceedings of the 36th
  International Conference on Machine Learning}}
  \emph{(\bibinfo{series}{Proceedings of Machine Learning Research},
  Vol.~\bibinfo{volume}{97})}, \bibfield{editor}{\bibinfo{person}{Kamalika
  Chaudhuri} {and} \bibinfo{person}{Ruslan Salakhutdinov}} (Eds.).
  \bibinfo{publisher}{PMLR}, \bibinfo{pages}{242--252}.
\newblock
\urldef\tempurl%
\url{https://proceedings.mlr.press/v97/allen-zhu19a.html}
\showURL{%
\tempurl}


\bibitem[Ashraf et~al\mbox{.}(2024)]%
        {ashraf2024castlecsf}
\bibfield{author}{\bibinfo{person}{Maliha Ashraf}, \bibinfo{person}{Rafa{\l}~K
  Mantiuk}, \bibinfo{person}{Alexandre Chapiro}, {and} \bibinfo{person}{Sophie
  Wuerger}.} \bibinfo{year}{2024}\natexlab{}.
\newblock \showarticletitle{castleCSF—A contrast sensitivity function of
  color, area, spatiotemporal frequency, luminance and eccentricity}.
\newblock \bibinfo{journal}{\emph{Journal of vision}} \bibinfo{volume}{24},
  \bibinfo{number}{4} (\bibinfo{year}{2024}), \bibinfo{pages}{5--5}.
\newblock


\bibitem[Chakravarthula et~al\mbox{.}(2022a)]%
        {chakravarthula2022pupil}
\bibfield{author}{\bibinfo{person}{Praneeth Chakravarthula},
  \bibinfo{person}{Seung-Hwan Baek}, \bibinfo{person}{Florian Schiffers},
  \bibinfo{person}{Ethan Tseng}, \bibinfo{person}{Grace Kuo},
  \bibinfo{person}{Andrew Maimone}, \bibinfo{person}{Nathan Matsuda},
  \bibinfo{person}{Oliver Cossairt}, \bibinfo{person}{Douglas Lanman}, {and}
  \bibinfo{person}{Felix Heide}.} \bibinfo{year}{2022}\natexlab{a}.
\newblock \showarticletitle{Pupil-Aware Holography}.
\newblock \bibinfo{journal}{\emph{ACM Trans. Graph.}} \bibinfo{volume}{41},
  \bibinfo{number}{6}, Article \bibinfo{articleno}{212} (\bibinfo{date}{Nov.}
  \bibinfo{year}{2022}), \bibinfo{numpages}{15}~pages.
\newblock
\showISSN{0730-0301}
\href{https://doi.org/10.1145/3550454.3555508}{doi:\nolinkurl{10.1145/3550454.3555508}}


\bibitem[Chakravarthula et~al\mbox{.}(2019)]%
        {chakravarthula2019wirtinger}
\bibfield{author}{\bibinfo{person}{Praneeth Chakravarthula},
  \bibinfo{person}{Yifan Peng}, \bibinfo{person}{Joel Kollin},
  \bibinfo{person}{Henry Fuchs}, {and} \bibinfo{person}{Felix Heide}.}
  \bibinfo{year}{2019}\natexlab{}.
\newblock \showarticletitle{Wirtinger holography for near-eye displays}.
\newblock \bibinfo{journal}{\emph{ACM Transactions on Graphics (TOG)}}
  \bibinfo{volume}{38}, \bibinfo{number}{6} (\bibinfo{year}{2019}),
  \bibinfo{pages}{1--13}.
\newblock


\bibitem[Chakravarthula et~al\mbox{.}(2022b)]%
        {chakravarthula2022differentiable}
\bibfield{author}{\bibinfo{person}{Praneeth Chakravarthula},
  \bibinfo{person}{Florian Schiffers}, \bibinfo{person}{Oliver Cossairt},
  \bibinfo{person}{Ethan Tseng}, \bibinfo{person}{Seung-Hwan Baek}, {and}
  \bibinfo{person}{Felix Heide}.} \bibinfo{year}{2022}\natexlab{b}.
\newblock \showarticletitle{Differentiable cameras and displays}. In
  \bibinfo{booktitle}{\emph{ACM SIGGRAPH 2022 Courses}} (Vancouver, British
  Columbia, Canada) \emph{(\bibinfo{series}{SIGGRAPH '22})}.
  \bibinfo{publisher}{Association for Computing Machinery},
  \bibinfo{address}{New York, NY, USA}, Article \bibinfo{articleno}{6},
  \bibinfo{numpages}{213}~pages.
\newblock
\showISBNx{9781450393621}
\href{https://doi.org/10.1145/3532720.3535685}{doi:\nolinkurl{10.1145/3532720.3535685}}


\bibitem[Chakravarthula et~al\mbox{.}(2022c)]%
        {chakravarthula2022hogel}
\bibfield{author}{\bibinfo{person}{Praneeth Chakravarthula},
  \bibinfo{person}{Ethan Tseng}, \bibinfo{person}{Henry Fuchs}, {and}
  \bibinfo{person}{Felix Heide}.} \bibinfo{year}{2022}\natexlab{c}.
\newblock \showarticletitle{Hogel-free holography}.
\newblock \bibinfo{journal}{\emph{ACM Transactions on Graphics}}
  \bibinfo{volume}{41}, \bibinfo{number}{5} (\bibinfo{year}{2022}),
  \bibinfo{pages}{1--16}.
\newblock


\bibitem[Chakravarthula et~al\mbox{.}(2020)]%
        {lut}
\bibfield{author}{\bibinfo{person}{Praneeth Chakravarthula},
  \bibinfo{person}{Ethan Tseng}, \bibinfo{person}{Tarun Srivastava},
  \bibinfo{person}{Henry Fuchs}, {and} \bibinfo{person}{Felix Heide}.}
  \bibinfo{year}{2020}\natexlab{}.
\newblock \showarticletitle{Learned hardware-in-the-loop phase retrieval for
  holographic near-eye displays}.
\newblock \bibinfo{journal}{\emph{ACM Trans. Graph.}} \bibinfo{volume}{39},
  \bibinfo{number}{6}, Article \bibinfo{articleno}{186} (\bibinfo{date}{Nov.}
  \bibinfo{year}{2020}), \bibinfo{numpages}{18}~pages.
\newblock
\showISSN{0730-0301}
\href{https://doi.org/10.1145/3414685.3417846}{doi:\nolinkurl{10.1145/3414685.3417846}}


\bibitem[Chakravarthula et~al\mbox{.}(2021)]%
        {chakravarthula2021gaze}
\bibfield{author}{\bibinfo{person}{Praneeth Chakravarthula},
  \bibinfo{person}{Zhan Zhang}, \bibinfo{person}{Okan Tursun},
  \bibinfo{person}{Piotr Didyk}, \bibinfo{person}{Qi Sun}, {and}
  \bibinfo{person}{Henry Fuchs}.} \bibinfo{year}{2021}\natexlab{}.
\newblock \showarticletitle{Gaze-contingent retinal speckle suppression for
  perceptually-matched foveated holographic displays}.
\newblock \bibinfo{journal}{\emph{IEEE Transactions on Visualization and
  Computer Graphics}} \bibinfo{volume}{27}, \bibinfo{number}{11}
  (\bibinfo{year}{2021}), \bibinfo{pages}{4194--4203}.
\newblock


\bibitem[Chapiro et~al\mbox{.}(2024)]%
        {xrdavid}
\bibfield{author}{\bibinfo{person}{Alexandre Chapiro}, \bibinfo{person}{Yuta
  Asano}, \bibinfo{person}{Rafal Mantiuk}, \bibinfo{person}{Param Hanji}, {and}
  \bibinfo{person}{Maliha Ashraf}.} \bibinfo{year}{2024}\natexlab{}.
\newblock \showarticletitle{XR-DAVID: XR Display Artifact Video Dataset}.
\newblock  (\bibinfo{year}{2024}).
\newblock
\href{https://doi.org/10.17863/CAM.108210}{doi:\nolinkurl{10.17863/CAM.108210}}


\bibitem[Chen et~al\mbox{.}(2024)]%
        {chen2024pea}
\bibfield{author}{\bibinfo{person}{Kenneth Chen}, \bibinfo{person}{Thomas Wan},
  \bibinfo{person}{Nathan Matsuda}, \bibinfo{person}{Ajit Ninan},
  \bibinfo{person}{Alexandre Chapiro}, {and} \bibinfo{person}{Qi Sun}.}
  \bibinfo{year}{2024}\natexlab{}.
\newblock \showarticletitle{Pea-pods: Perceptual evaluation of algorithms for
  power optimization in xr displays}.
\newblock \bibinfo{journal}{\emph{ACM Transactions on Graphics (TOG)}}
  \bibinfo{volume}{43}, \bibinfo{number}{4} (\bibinfo{year}{2024}),
  \bibinfo{pages}{1--17}.
\newblock


\bibitem[Choi et~al\mbox{.}(2025)]%
        {choi_gaussian_2025}
\bibfield{author}{\bibinfo{person}{Suyeon Choi}, \bibinfo{person}{Brian Chao},
  \bibinfo{person}{Jacqueline Yang}, \bibinfo{person}{Manu Gopakumar}, {and}
  \bibinfo{person}{Gordon Wetzstein}.} \bibinfo{year}{2025}\natexlab{}.
\newblock \showarticletitle{Gaussian Wave Splatting for Computer-Generated
  Holography}.
\newblock \bibinfo{journal}{\emph{ACM Trans. Graph.}} \bibinfo{volume}{44},
  \bibinfo{number}{4}, Article \bibinfo{articleno}{57} (\bibinfo{date}{July}
  \bibinfo{year}{2025}), \bibinfo{numpages}{13}~pages.
\newblock
\showISSN{0730-0301}
\href{https://doi.org/10.1145/3731163}{doi:\nolinkurl{10.1145/3731163}}


\bibitem[Choi et~al\mbox{.}(2022)]%
        {choi2022time}
\bibfield{author}{\bibinfo{person}{Suyeon Choi}, \bibinfo{person}{Manu
  Gopakumar}, \bibinfo{person}{Yifan Peng}, \bibinfo{person}{Jonghyun Kim},
  \bibinfo{person}{Matthew O'Toole}, {and} \bibinfo{person}{Gordon Wetzstein}.}
  \bibinfo{year}{2022}\natexlab{}.
\newblock \showarticletitle{Time-multiplexed Neural Holography: A Flexible
  Framework for Holographic Near-eye Displays with Fast Heavily-quantized
  Spatial Light Modulators}. In \bibinfo{booktitle}{\emph{ACM SIGGRAPH 2022
  Conference Proceedings}} (Vancouver, BC, Canada)
  \emph{(\bibinfo{series}{SIGGRAPH '22})}. \bibinfo{publisher}{Association for
  Computing Machinery}, \bibinfo{address}{New York, NY, USA}, Article
  \bibinfo{articleno}{32}, \bibinfo{numpages}{9}~pages.
\newblock
\showISBNx{9781450393379}
\href{https://doi.org/10.1145/3528233.3530734}{doi:\nolinkurl{10.1145/3528233.3530734}}


\bibitem[Cogalan et~al\mbox{.}(2025)]%
        {cogalan2025milo}
\bibfield{author}{\bibinfo{person}{Ugur Cogalan}, \bibinfo{person}{Mojtaba
  Bemana}, \bibinfo{person}{Karol Myszkowski}, \bibinfo{person}{Hans-Peter
  Seidel}, {and} \bibinfo{person}{Colin Groth}.}
  \bibinfo{year}{2025}\natexlab{}.
\newblock \showarticletitle{MILO: A Lightweight Perceptual Quality Metric for
  Image and Latent-Space Optimization}.
\newblock \bibinfo{journal}{\emph{ACM Transactions on Graphics (TOG)}}
  \bibinfo{volume}{44}, \bibinfo{number}{6} (\bibinfo{year}{2025}),
  \bibinfo{pages}{1--11}.
\newblock


\bibitem[Ding et~al\mbox{.}(2020)]%
        {ding2020image}
\bibfield{author}{\bibinfo{person}{Keyan Ding}, \bibinfo{person}{Kede Ma},
  \bibinfo{person}{Shiqi Wang}, {and} \bibinfo{person}{Eero~P Simoncelli}.}
  \bibinfo{year}{2020}\natexlab{}.
\newblock \showarticletitle{Image quality assessment: Unifying structure and
  texture similarity}.
\newblock \bibinfo{journal}{\emph{IEEE transactions on pattern analysis and
  machine intelligence}} \bibinfo{volume}{44}, \bibinfo{number}{5}
  (\bibinfo{year}{2020}), \bibinfo{pages}{2567--2581}.
\newblock


\bibitem[Ding et~al\mbox{.}(2021)]%
        {ding2021comparison}
\bibfield{author}{\bibinfo{person}{Keyan Ding}, \bibinfo{person}{Kede Ma},
  \bibinfo{person}{Shiqi Wang}, {and} \bibinfo{person}{Eero~P Simoncelli}.}
  \bibinfo{year}{2021}\natexlab{}.
\newblock \showarticletitle{Comparison of full-reference image quality models
  for optimization of image processing systems}.
\newblock \bibinfo{journal}{\emph{International Journal of Computer Vision}}
  \bibinfo{volume}{129}, \bibinfo{number}{4} (\bibinfo{year}{2021}),
  \bibinfo{pages}{1258--1281}.
\newblock


\bibitem[Du et~al\mbox{.}(2019)]%
        {du2018gradient}
\bibfield{author}{\bibinfo{person}{Simon~S. Du}, \bibinfo{person}{Xiyu Zhai},
  \bibinfo{person}{Barnabas Poczos}, {and} \bibinfo{person}{Aarti Singh}.}
  \bibinfo{year}{2019}\natexlab{}.
\newblock \showarticletitle{Gradient Descent Provably Optimizes
  Over-parameterized Neural Networks}. In
  \bibinfo{booktitle}{\emph{International Conference on Learning
  Representations}}.
\newblock
\urldef\tempurl%
\url{https://openreview.net/forum?id=S1eK3i09YQ}
\showURL{%
\tempurl}


\bibitem[Freeman and Simoncelli(2011)]%
        {freeman2011metamers}
\bibfield{author}{\bibinfo{person}{Jeremy Freeman} {and}
  \bibinfo{person}{Eero~P Simoncelli}.} \bibinfo{year}{2011}\natexlab{}.
\newblock \showarticletitle{Metamers of the ventral stream}.
\newblock \bibinfo{journal}{\emph{Nature neuroscience}} \bibinfo{volume}{14},
  \bibinfo{number}{9} (\bibinfo{year}{2011}), \bibinfo{pages}{1195--1201}.
\newblock


\bibitem[Georgiou et~al\mbox{.}(2023)]%
        {georgiou2023visual}
\bibfield{author}{\bibinfo{person}{Andreas Georgiou}, \bibinfo{person}{Joel
  Kollin}, \bibinfo{person}{Charlie Hewitt}, \bibinfo{person}{Praneeth
  Chakravarthula}, {and} \bibinfo{person}{Brian Guenter}.}
  \bibinfo{year}{2023}\natexlab{}.
\newblock \showarticletitle{Visual perception of noise in a simulated
  holographic display—a user study}.
\newblock \bibinfo{journal}{\emph{Displays}}  \bibinfo{volume}{76}
  (\bibinfo{year}{2023}), \bibinfo{pages}{102333}.
\newblock


\bibitem[Goodman(2007)]%
        {goodman2007speckle}
\bibfield{author}{\bibinfo{person}{Joseph~W Goodman}.}
  \bibinfo{year}{2007}\natexlab{}.
\newblock \bibinfo{booktitle}{\emph{Speckle phenomena in optics: theory and
  applications}}.
\newblock \bibinfo{publisher}{Roberts and company Publishers}.
\newblock


\bibitem[{International Commission on Illumination}(2019)]%
        {lab}
\bibfield{author}{\bibinfo{person}{{International Commission on
  Illumination}}.} \bibinfo{year}{2019}\natexlab{}.
\newblock \bibinfo{title}{{ISO}/{CIE} 11664-4:2019}.
\newblock
\urldef\tempurl%
\url{https://www.iso.org/standard/74166.html}
\showURL{%
\tempurl}


\bibitem[{International Telecommunication Union}(2025)]%
        {ictcp}
\bibfield{author}{\bibinfo{person}{{International Telecommunication Union}}.}
  \bibinfo{year}{2025}\natexlab{}.
\newblock \bibinfo{title}{{BT}.2100 : {Image} parameter values for high dynamic
  range television for use in production and international programme exchange}.
\newblock
\urldef\tempurl%
\url{https://www.itu.int/rec/R-REC-BT.2100}
\showURL{%
\tempurl}


\bibitem[Iwai and Asakura(1996)]%
        {iwai1996speckle}
\bibfield{author}{\bibinfo{person}{Toshiaki Iwai} {and}
  \bibinfo{person}{Toshimitsu Asakura}.} \bibinfo{year}{1996}\natexlab{}.
\newblock \showarticletitle{Speckle reduction in coherent information
  processing}.
\newblock \bibinfo{journal}{\emph{Proc. IEEE}} \bibinfo{volume}{84},
  \bibinfo{number}{5} (\bibinfo{year}{1996}), \bibinfo{pages}{765--781}.
\newblock


\bibitem[Kirillov et~al\mbox{.}(2023)]%
        {kirillov2023segment}
\bibfield{author}{\bibinfo{person}{Alexander Kirillov}, \bibinfo{person}{Eric
  Mintun}, \bibinfo{person}{Nikhila Ravi}, \bibinfo{person}{Hanzi Mao},
  \bibinfo{person}{Chloe Rolland}, \bibinfo{person}{Laura Gustafson},
  \bibinfo{person}{Tete Xiao}, \bibinfo{person}{Spencer Whitehead},
  \bibinfo{person}{Alexander~C Berg}, \bibinfo{person}{Wan-Yen Lo},
  {et~al\mbox{.}}} \bibinfo{year}{2023}\natexlab{}.
\newblock \showarticletitle{Segment anything}. In
  \bibinfo{booktitle}{\emph{Proceedings of the IEEE/CVF international
  conference on computer vision}}. \bibinfo{pages}{4015--4026}.
\newblock


\bibitem[Krynski(2022)]%
        {krynski_blender_2022}
\bibfield{author}{\bibinfo{person}{Piotr Krynski}.}
  \bibinfo{year}{2022}\natexlab{}.
\newblock \bibinfo{title}{Blender 3.3 {LTS} - {Scanlands}}.
\newblock
\urldef\tempurl%
\url{https://www.blender.org/download/demo/splash/blender-3.3-splash.blend}
\showURL{%
\tempurl}


\bibitem[Laparra et~al\mbox{.}(2017)]%
        {laparra2017perceptually}
\bibfield{author}{\bibinfo{person}{Valero Laparra}, \bibinfo{person}{Alexander
  Berardino}, \bibinfo{person}{Johannes Ball{\'e}}, {and}
  \bibinfo{person}{Eero~P Simoncelli}.} \bibinfo{year}{2017}\natexlab{}.
\newblock \showarticletitle{Perceptually optimized image rendering}.
\newblock \bibinfo{journal}{\emph{Journal of the Optical Society of America A}}
  \bibinfo{volume}{34}, \bibinfo{number}{9} (\bibinfo{year}{2017}),
  \bibinfo{pages}{1511--1525}.
\newblock


\bibitem[Lee et~al\mbox{.}(2022)]%
        {lee2022high}
\bibfield{author}{\bibinfo{person}{Byounghyo Lee}, \bibinfo{person}{Dongyeon
  Kim}, \bibinfo{person}{Seungjae Lee}, \bibinfo{person}{Chun Chen}, {and}
  \bibinfo{person}{Byoungho Lee}.} \bibinfo{year}{2022}\natexlab{}.
\newblock \showarticletitle{High-contrast, speckle-free, true 3D holography via
  binary CGH optimization}.
\newblock \bibinfo{journal}{\emph{Scientific reports}} \bibinfo{volume}{12},
  \bibinfo{number}{1} (\bibinfo{year}{2022}), \bibinfo{pages}{2811}.
\newblock


\bibitem[Li et~al\mbox{.}(2017)]%
        {cam16}
\bibfield{author}{\bibinfo{person}{Changjun Li}, \bibinfo{person}{Zhiqiang Li},
  \bibinfo{person}{Zhifeng Wang}, \bibinfo{person}{Yang Xu},
  \bibinfo{person}{Ming~Ronnier Luo}, \bibinfo{person}{Guihua Cui},
  \bibinfo{person}{Manuel Melgosa}, \bibinfo{person}{Michael~H Brill}, {and}
  \bibinfo{person}{Michael Pointer}.} \bibinfo{year}{2017}\natexlab{}.
\newblock \showarticletitle{Comprehensive color solutions: CAM16, CAT16, and
  CAM16-UCS}.
\newblock \bibinfo{journal}{\emph{Color Research \& Application}}
  \bibinfo{volume}{42}, \bibinfo{number}{6} (\bibinfo{year}{2017}),
  \bibinfo{pages}{703--718}.
\newblock


\bibitem[Liu et~al\mbox{.}(2025)]%
        {liu2025propagation}
\bibfield{author}{\bibinfo{person}{Ninghe Liu}, \bibinfo{person}{Kexuan Liu},
  \bibinfo{person}{Yixin Yang}, \bibinfo{person}{Yifan Peng}, {and}
  \bibinfo{person}{Liangcai Cao}.} \bibinfo{year}{2025}\natexlab{}.
\newblock \showarticletitle{Propagation-adaptive 4K computer-generated
  holography using physics-constrained spatial and Fourier neural operator}.
\newblock \bibinfo{journal}{\emph{Nature communications}} \bibinfo{volume}{16},
  \bibinfo{number}{1} (\bibinfo{year}{2025}), \bibinfo{pages}{7761}.
\newblock


\bibitem[Liu and Takaki(2020)]%
        {liu2020optimization}
\bibfield{author}{\bibinfo{person}{Shujian Liu} {and} \bibinfo{person}{Yasuhiro
  Takaki}.} \bibinfo{year}{2020}\natexlab{}.
\newblock \showarticletitle{Optimization of phase-only computer-generated
  holograms based on the gradient descent method}.
\newblock \bibinfo{journal}{\emph{Applied Sciences}} \bibinfo{volume}{10},
  \bibinfo{number}{12} (\bibinfo{year}{2020}), \bibinfo{pages}{4283}.
\newblock


\bibitem[Maclaurin et~al\mbox{.}(2015)]%
        {hyperparameter}
\bibfield{author}{\bibinfo{person}{Dougal Maclaurin}, \bibinfo{person}{David
  Duvenaud}, {and} \bibinfo{person}{Ryan Adams}.}
  \bibinfo{year}{2015}\natexlab{}.
\newblock \showarticletitle{Gradient-based hyperparameter optimization through
  reversible learning}. In \bibinfo{booktitle}{\emph{International conference
  on machine learning}}. PMLR, \bibinfo{pages}{2113--2122}.
\newblock


\bibitem[Maimone et~al\mbox{.}(2017)]%
        {maimone2017holographic}
\bibfield{author}{\bibinfo{person}{Andrew Maimone}, \bibinfo{person}{Andreas
  Georgiou}, {and} \bibinfo{person}{Joel~S Kollin}.}
  \bibinfo{year}{2017}\natexlab{}.
\newblock \showarticletitle{Holographic near-eye displays for virtual and
  augmented reality}.
\newblock \bibinfo{journal}{\emph{ACM Transactions on Graphics (Tog)}}
  \bibinfo{volume}{36}, \bibinfo{number}{4} (\bibinfo{year}{2017}),
  \bibinfo{pages}{1--16}.
\newblock


\bibitem[Mantiuk et~al\mbox{.}(2021)]%
        {mantiuk2021fovvideovdp}
\bibfield{author}{\bibinfo{person}{Rafa{\l}~K Mantiuk}, \bibinfo{person}{Gyorgy
  Denes}, \bibinfo{person}{Alexandre Chapiro}, \bibinfo{person}{Anton
  Kaplanyan}, \bibinfo{person}{Gizem Rufo}, \bibinfo{person}{Romain Bachy},
  \bibinfo{person}{Trisha Lian}, {and} \bibinfo{person}{Anjul Patney}.}
  \bibinfo{year}{2021}\natexlab{}.
\newblock \showarticletitle{Fovvideovdp: A visible difference predictor for
  wide field-of-view video}.
\newblock \bibinfo{journal}{\emph{ACM Transactions on Graphics (TOG)}}
  \bibinfo{volume}{40}, \bibinfo{number}{4} (\bibinfo{year}{2021}),
  \bibinfo{pages}{1--19}.
\newblock


\bibitem[Mantiuk et~al\mbox{.}(2024)]%
        {cvvdp}
\bibfield{author}{\bibinfo{person}{Rafal~K. Mantiuk}, \bibinfo{person}{Param
  Hanji}, \bibinfo{person}{Maliha Ashraf}, \bibinfo{person}{Yuta Asano}, {and}
  \bibinfo{person}{Alexandre Chapiro}.} \bibinfo{year}{2024}\natexlab{}.
\newblock \showarticletitle{ColorVideoVDP: A visual difference predictor for
  image, video and display distortions}.
\newblock \bibinfo{journal}{\emph{ACM Trans. Graph.}} \bibinfo{volume}{43},
  \bibinfo{number}{4}, Article \bibinfo{articleno}{129} (\bibinfo{date}{July}
  \bibinfo{year}{2024}), \bibinfo{numpages}{20}~pages.
\newblock
\showISSN{0730-0301}
\href{https://doi.org/10.1145/3658144}{doi:\nolinkurl{10.1145/3658144}}


\bibitem[Markley et~al\mbox{.}(2023)]%
        {sccgh}
\bibfield{author}{\bibinfo{person}{Eric Markley}, \bibinfo{person}{Nathan
  Matsuda}, \bibinfo{person}{Florian Schiffers}, \bibinfo{person}{Oliver
  Cossairt}, {and} \bibinfo{person}{Grace Kuo}.}
  \bibinfo{year}{2023}\natexlab{}.
\newblock \showarticletitle{Simultaneous Color Computer Generated Holography}.
  In \bibinfo{booktitle}{\emph{SIGGRAPH Asia 2023 Conference Papers}} (Sydney,
  NSW, Australia) \emph{(\bibinfo{series}{SA '23})}.
  \bibinfo{publisher}{Association for Computing Machinery},
  \bibinfo{address}{New York, NY, USA}, Article \bibinfo{articleno}{22},
  \bibinfo{numpages}{11}~pages.
\newblock
\showISBNx{9798400703157}
\href{https://doi.org/10.1145/3610548.3618250}{doi:\nolinkurl{10.1145/3610548.3618250}}


\bibitem[Matsushima and Shimobaba(2009)]%
        {matsushima2009band}
\bibfield{author}{\bibinfo{person}{Kyoji Matsushima} {and}
  \bibinfo{person}{Tomoyoshi Shimobaba}.} \bibinfo{year}{2009}\natexlab{}.
\newblock \showarticletitle{Band-limited angular spectrum method for numerical
  simulation of free-space propagation in far and near fields}.
\newblock \bibinfo{journal}{\emph{Optics express}} \bibinfo{volume}{17},
  \bibinfo{number}{22} (\bibinfo{year}{2009}), \bibinfo{pages}{19662--19673}.
\newblock


\bibitem[Mikhailiuk et~al\mbox{.}(2021)]%
        {asap}
\bibfield{author}{\bibinfo{person}{Aliaksei Mikhailiuk},
  \bibinfo{person}{Clifford Wilmot}, \bibinfo{person}{Maria Perez-Ortiz},
  \bibinfo{person}{Dingcheng Yue}, {and} \bibinfo{person}{Rafal Mantiuk}.}
  \bibinfo{year}{2021}\natexlab{}.
\newblock \showarticletitle{Active Sampling for Pairwise Comparisons via
  Approximate Message Passing and Information Gain Maximization}. In
  \bibinfo{booktitle}{\emph{2020 IEEE International Conference on Pattern
  Recognition (ICPR)}}.
\newblock


\bibitem[Morena(2023)]%
        {morena_blender_2023}
\bibfield{author}{\bibinfo{person}{Nicole Morena}.}
  \bibinfo{year}{2023}\natexlab{}.
\newblock \bibinfo{title}{Blender 3.5 - {Cozy} {Kitchen}}.
\newblock
\urldef\tempurl%
\url{https://www.blender.org/download/demo/splash/blender-3.5-splash.blend}
\showURL{%
\tempurl}


\bibitem[Mori et~al\mbox{.}(2014)]%
        {Mori:14}
\bibfield{author}{\bibinfo{person}{Yutaka Mori}, \bibinfo{person}{Takahiko
  Fukuoka}, {and} \bibinfo{person}{Takanori Nomura}.}
  \bibinfo{year}{2014}\natexlab{}.
\newblock \showarticletitle{Speckle reduction in holographic projection by
  random pixel separation with time multiplexing}.
\newblock \bibinfo{journal}{\emph{Appl. Opt.}} \bibinfo{volume}{53},
  \bibinfo{number}{35} (\bibinfo{date}{Dec} \bibinfo{year}{2014}),
  \bibinfo{pages}{8182--8188}.
\newblock
\href{https://doi.org/10.1364/AO.53.008182}{doi:\nolinkurl{10.1364/AO.53.008182}}


\bibitem[Mustafa et~al\mbox{.}(2022)]%
        {mustafa2022training}
\bibfield{author}{\bibinfo{person}{Aamir Mustafa}, \bibinfo{person}{Aliaksei
  Mikhailiuk}, \bibinfo{person}{Dan~Andrei Iliescu}, \bibinfo{person}{Varun
  Babbar}, {and} \bibinfo{person}{Rafa{\l}~K Mantiuk}.}
  \bibinfo{year}{2022}\natexlab{}.
\newblock \showarticletitle{Training a task-specific image reconstruction
  loss}. In \bibinfo{booktitle}{\emph{Proceedings of the IEEE/CVF winter
  conference on applications of computer vision}}. \bibinfo{pages}{2319--2328}.
\newblock


\bibitem[Nomura and Shibata(2024)]%
        {cmaes}
\bibfield{author}{\bibinfo{person}{Masahiro Nomura} {and}
  \bibinfo{person}{Masashi Shibata}.} \bibinfo{year}{2024}\natexlab{}.
\newblock \showarticletitle{cmaes : A Simple yet Practical Python Library for
  CMA-ES}.
\newblock \bibinfo{journal}{\emph{arXiv preprint arXiv:2402.01373}}
  (\bibinfo{year}{2024}).
\newblock


\bibitem[Ottosson(2020)]%
        {oklab}
\bibfield{author}{\bibinfo{person}{Björn Ottosson}.}
  \bibinfo{year}{2020}\natexlab{}.
\newblock \bibinfo{title}{A perceptual color space for image processing}.
\newblock
\urldef\tempurl%
\url{https://bottosson.github.io/posts/oklab/}
\showURL{%
\tempurl}


\bibitem[Padmanaban et~al\mbox{.}(2019)]%
        {padmanaban2019holographic}
\bibfield{author}{\bibinfo{person}{Nitish Padmanaban}, \bibinfo{person}{Yifan
  Peng}, {and} \bibinfo{person}{Gordon Wetzstein}.}
  \bibinfo{year}{2019}\natexlab{}.
\newblock \showarticletitle{Holographic near-eye displays based on overlap-add
  stereograms}.
\newblock \bibinfo{journal}{\emph{ACM Transactions on Graphics (TOG)}}
  \bibinfo{volume}{38}, \bibinfo{number}{6} (\bibinfo{year}{2019}),
  \bibinfo{pages}{1--13}.
\newblock


\bibitem[Peli(1990)]%
        {peli1990contrast}
\bibfield{author}{\bibinfo{person}{Eli Peli}.} \bibinfo{year}{1990}\natexlab{}.
\newblock \showarticletitle{Contrast in complex images}.
\newblock \bibinfo{journal}{\emph{Journal of the Optical Society of America A}}
  \bibinfo{volume}{7}, \bibinfo{number}{10} (\bibinfo{year}{1990}),
  \bibinfo{pages}{2032--2040}.
\newblock


\bibitem[Peng et~al\mbox{.}(2020)]%
        {peng2020neural}
\bibfield{author}{\bibinfo{person}{Yifan Peng}, \bibinfo{person}{Suyeon Choi},
  \bibinfo{person}{Nitish Padmanaban}, {and} \bibinfo{person}{Gordon
  Wetzstein}.} \bibinfo{year}{2020}\natexlab{}.
\newblock \showarticletitle{Neural holography with camera-in-the-loop
  training}.
\newblock \bibinfo{journal}{\emph{ACM Transactions on Graphics (TOG)}}
  \bibinfo{volume}{39}, \bibinfo{number}{6} (\bibinfo{year}{2020}),
  \bibinfo{pages}{1--14}.
\newblock


\bibitem[Perez-Ortiz and Mantiuk(2017)]%
        {pwcmp}
\bibfield{author}{\bibinfo{person}{Maria Perez-Ortiz} {and}
  \bibinfo{person}{Rafal~K Mantiuk}.} \bibinfo{year}{2017}\natexlab{}.
\newblock \showarticletitle{A practical guide and software for analysing
  pairwise comparison experiments}.
\newblock \bibinfo{journal}{\emph{arXiv preprint arXiv:1712.03686}}
  (\bibinfo{year}{2017}).
\newblock


\bibitem[Ponomarenko et~al\mbox{.}(2015)]%
        {ponomarenko2015image}
\bibfield{author}{\bibinfo{person}{Nikolay Ponomarenko}, \bibinfo{person}{Lina
  Jin}, \bibinfo{person}{Oleg Ieremeiev}, \bibinfo{person}{Vladimir Lukin},
  \bibinfo{person}{Karen Egiazarian}, \bibinfo{person}{Jaakko Astola},
  \bibinfo{person}{Benoit Vozel}, \bibinfo{person}{Kacem Chehdi},
  \bibinfo{person}{Marco Carli}, \bibinfo{person}{Federica Battisti},
  {et~al\mbox{.}}} \bibinfo{year}{2015}\natexlab{}.
\newblock \showarticletitle{Image database TID2013: Peculiarities, results and
  perspectives}.
\newblock \bibinfo{journal}{\emph{Signal processing: Image communication}}
  \bibinfo{volume}{30} (\bibinfo{year}{2015}), \bibinfo{pages}{57--77}.
\newblock


\bibitem[Ram(2018)]%
        {ram_eurasian_2018}
\bibfield{author}{\bibinfo{person}{M~S Ram}.} \bibinfo{year}{2018}\natexlab{}.
\newblock \bibinfo{title}{Eurasian hoopoe ({Upupa} epops) dust-bathing}.
\newblock
\urldef\tempurl%
\url{https://www.youtube.com/watch?v=xm93eB8aXHE}
\showURL{%
\tempurl}


\bibitem[Robson(1966)]%
        {Robson:66}
\bibfield{author}{\bibinfo{person}{J.~G. Robson}.}
  \bibinfo{year}{1966}\natexlab{}.
\newblock \showarticletitle{Spatial and Temporal Contrast-Sensitivity Functions
  of the Visual System}.
\newblock \bibinfo{journal}{\emph{J. Opt. Soc. Am.}} \bibinfo{volume}{56},
  \bibinfo{number}{8} (\bibinfo{date}{Aug} \bibinfo{year}{1966}),
  \bibinfo{pages}{1141--1142}.
\newblock
\href{https://doi.org/10.1364/JOSA.56.001141}{doi:\nolinkurl{10.1364/JOSA.56.001141}}


\bibitem[Schiffers et~al\mbox{.}(2023)]%
        {schiffers2023stochastic}
\bibfield{author}{\bibinfo{person}{Florian Schiffers},
  \bibinfo{person}{Praneeth Chakravarthula}, \bibinfo{person}{Nathan Matsuda},
  \bibinfo{person}{Grace Kuo}, \bibinfo{person}{Ethan Tseng},
  \bibinfo{person}{Douglas Lanman}, \bibinfo{person}{Felix Heide}, {and}
  \bibinfo{person}{Oliver Cossairt}.} \bibinfo{year}{2023}\natexlab{}.
\newblock \showarticletitle{Stochastic light field holography}. In
  \bibinfo{booktitle}{\emph{2023 IEEE International Conference on Computational
  Photography (ICCP)}}. IEEE, \bibinfo{pages}{1--12}.
\newblock


\bibitem[Sha et~al\mbox{.}(2024)]%
        {sha2024multi}
\bibfield{author}{\bibinfo{person}{Jinze Sha}, \bibinfo{person}{Antoni Wojcik},
  \bibinfo{person}{Benjamin Wetherfield}, \bibinfo{person}{Jianghan Yu}, {and}
  \bibinfo{person}{Timothy~D Wilkinson}.} \bibinfo{year}{2024}\natexlab{}.
\newblock \showarticletitle{Multi frame holograms batched optimization for
  binary phase spatial light modulators}.
\newblock \bibinfo{journal}{\emph{Scientific Reports}} \bibinfo{volume}{14},
  \bibinfo{number}{1} (\bibinfo{year}{2024}), \bibinfo{pages}{19380}.
\newblock


\bibitem[Shi et~al\mbox{.}(2021)]%
        {shi2021towards}
\bibfield{author}{\bibinfo{person}{Liang Shi}, \bibinfo{person}{Beichen Li},
  \bibinfo{person}{Changil Kim}, \bibinfo{person}{Petr Kellnhofer}, {and}
  \bibinfo{person}{Wojciech Matusik}.} \bibinfo{year}{2021}\natexlab{}.
\newblock \showarticletitle{Towards real-time photorealistic 3D holography with
  deep neural networks}.
\newblock \bibinfo{journal}{\emph{Nature}} \bibinfo{volume}{591},
  \bibinfo{number}{7849} (\bibinfo{year}{2021}), \bibinfo{pages}{234--239}.
\newblock


\bibitem[Tada(2023)]%
        {tada_blender_2023}
\bibfield{author}{\bibinfo{person}{Gaku Tada}.}
  \bibinfo{year}{2023}\natexlab{}.
\newblock \bibinfo{title}{Blender 4.0 - {Gaku}}.
\newblock
\urldef\tempurl%
\url{https://www.blender.org/download/demo/splash/blender-4.0-splash.blend}
\showURL{%
\tempurl}


\bibitem[Takaki and Yokouchi(2011)]%
        {takaki2011speckle}
\bibfield{author}{\bibinfo{person}{Yasuhiro Takaki} {and}
  \bibinfo{person}{Masahito Yokouchi}.} \bibinfo{year}{2011}\natexlab{}.
\newblock \showarticletitle{Speckle-free and grayscale hologram reconstruction
  using time-multiplexing technique}.
\newblock \bibinfo{journal}{\emph{Optics express}} \bibinfo{volume}{19},
  \bibinfo{number}{8} (\bibinfo{year}{2011}), \bibinfo{pages}{7567--7579}.
\newblock


\bibitem[Tariq et~al\mbox{.}(2023)]%
        {tariq2023perceptually}
\bibfield{author}{\bibinfo{person}{Taimoor Tariq}, \bibinfo{person}{Nathan
  Matsuda}, \bibinfo{person}{Eric Penner}, \bibinfo{person}{Jerry Jia},
  \bibinfo{person}{Douglas Lanman}, \bibinfo{person}{Ajit Ninan}, {and}
  \bibinfo{person}{Alexandre Chapiro}.} \bibinfo{year}{2023}\natexlab{}.
\newblock \showarticletitle{Perceptually adaptive real-time tone mapping}. In
  \bibinfo{booktitle}{\emph{SIGGRAPH Asia 2023 Conference Papers}}.
  \bibinfo{pages}{1--10}.
\newblock


\bibitem[Tursun et~al\mbox{.}(2019)]%
        {tursun2019luminance}
\bibfield{author}{\bibinfo{person}{Okan~Tarhan Tursun}, \bibinfo{person}{Elena
  Arabadzhiyska-Koleva}, \bibinfo{person}{Marek Wernikowski},
  \bibinfo{person}{Rados{\l}aw Mantiuk}, \bibinfo{person}{Hans-Peter Seidel},
  \bibinfo{person}{Karol Myszkowski}, {and} \bibinfo{person}{Piotr Didyk}.}
  \bibinfo{year}{2019}\natexlab{}.
\newblock \showarticletitle{Luminance-contrast-aware foveated rendering}.
\newblock \bibinfo{journal}{\emph{ACM Transactions on Graphics (TOG)}}
  \bibinfo{volume}{38}, \bibinfo{number}{4} (\bibinfo{year}{2019}),
  \bibinfo{pages}{1--14}.
\newblock


\bibitem[Walton et~al\mbox{.}(2021)]%
        {walton2021beyond}
\bibfield{author}{\bibinfo{person}{David~R Walton},
  \bibinfo{person}{Rafael~Kuffner Dos~Anjos}, \bibinfo{person}{Sebastian
  Friston}, \bibinfo{person}{David Swapp}, \bibinfo{person}{Kaan Ak{\c{s}}it},
  \bibinfo{person}{Anthony Steed}, {and} \bibinfo{person}{Tobias Ritschel}.}
  \bibinfo{year}{2021}\natexlab{}.
\newblock \showarticletitle{Beyond blur: Real-time ventral metamers for
  foveated rendering}.
\newblock \bibinfo{journal}{\emph{ACM Transactions on Graphics}}
  \bibinfo{volume}{40}, \bibinfo{number}{4} (\bibinfo{year}{2021}),
  \bibinfo{pages}{1--14}.
\newblock


\bibitem[Walton et~al\mbox{.}(2022)]%
        {walton2022metameric}
\bibfield{author}{\bibinfo{person}{David~R Walton}, \bibinfo{person}{Koray
  Kavakl{\i}}, \bibinfo{person}{Rafael~Kuffner Dos~Anjos},
  \bibinfo{person}{David Swapp}, \bibinfo{person}{Tim Weyrich},
  \bibinfo{person}{Hakan Urey}, \bibinfo{person}{Anthony Steed},
  \bibinfo{person}{Tobias Ritschel}, {and} \bibinfo{person}{Kaan Ak{\c{s}}it}.}
  \bibinfo{year}{2022}\natexlab{}.
\newblock \showarticletitle{Metameric varifocal holograms}. In
  \bibinfo{booktitle}{\emph{2022 IEEE Conference on Virtual Reality and 3D User
  Interfaces (VR)}}. IEEE, \bibinfo{pages}{746--755}.
\newblock


\bibitem[Wang et~al\mbox{.}(2004)]%
        {ssim}
\bibfield{author}{\bibinfo{person}{Zhou Wang}, \bibinfo{person}{A.C. Bovik},
  \bibinfo{person}{H.R. Sheikh}, {and} \bibinfo{person}{E.P. Simoncelli}.}
  \bibinfo{year}{2004}\natexlab{}.
\newblock \showarticletitle{Image quality assessment: from error visibility to
  structural similarity}.
\newblock \bibinfo{journal}{\emph{IEEE Transactions on Image Processing}}
  \bibinfo{volume}{13}, \bibinfo{number}{4} (\bibinfo{year}{2004}),
  \bibinfo{pages}{600--612}.
\newblock
\href{https://doi.org/10.1109/TIP.2003.819861}{doi:\nolinkurl{10.1109/TIP.2003.819861}}


\bibitem[Wang et~al\mbox{.}(2003)]%
        {msssim}
\bibfield{author}{\bibinfo{person}{Z. Wang}, \bibinfo{person}{E.P. Simoncelli},
  {and} \bibinfo{person}{A.C. Bovik}.} \bibinfo{year}{2003}\natexlab{}.
\newblock \showarticletitle{Multiscale structural similarity for image quality
  assessment}. In \bibinfo{booktitle}{\emph{The Thrity-Seventh Asilomar
  Conference on Signals, Systems \& Computers, 2003}},
  Vol.~\bibinfo{volume}{2}. \bibinfo{pages}{1398--1402 Vol.2}.
\newblock
\href{https://doi.org/10.1109/ACSSC.2003.1292216}{doi:\nolinkurl{10.1109/ACSSC.2003.1292216}}


\bibitem[Yang et~al\mbox{.}(2022)]%
        {yang2022perceptually}
\bibfield{author}{\bibinfo{person}{Fan Yang}, \bibinfo{person}{Andrew Kadis},
  \bibinfo{person}{Ralf Mouthaan}, \bibinfo{person}{Benjamin Wetherfield},
  \bibinfo{person}{Andrzej Kaczorowski}, {and} \bibinfo{person}{Timothy~D
  Wilkinson}.} \bibinfo{year}{2022}\natexlab{}.
\newblock \showarticletitle{Perceptually motivated loss functions for computer
  generated holographic displays}.
\newblock \bibinfo{journal}{\emph{Scientific reports}} \bibinfo{volume}{12},
  \bibinfo{number}{1} (\bibinfo{year}{2022}), \bibinfo{pages}{7709}.
\newblock


\bibitem[Zhang et~al\mbox{.}(2017)]%
        {zhang20173d}
\bibfield{author}{\bibinfo{person}{Jingzhao Zhang}, \bibinfo{person}{Nicolas
  P{\'e}gard}, \bibinfo{person}{Jingshan Zhong}, \bibinfo{person}{Hillel
  Adesnik}, {and} \bibinfo{person}{Laura Waller}.}
  \bibinfo{year}{2017}\natexlab{}.
\newblock \showarticletitle{3D computer-generated holography by non-convex
  optimization}.
\newblock \bibinfo{journal}{\emph{Optica}} \bibinfo{volume}{4},
  \bibinfo{number}{10} (\bibinfo{year}{2017}), \bibinfo{pages}{1306--1313}.
\newblock


\bibitem[Zhang et~al\mbox{.}(2018)]%
        {zhang2018perceptual}
\bibfield{author}{\bibinfo{person}{Richard Zhang}, \bibinfo{person}{Phillip
  Isola}, \bibinfo{person}{Alexei~A. Efros}, \bibinfo{person}{Eli Shechtman},
  {and} \bibinfo{person}{Oliver Wang}.} \bibinfo{year}{2018}\natexlab{}.
\newblock \showarticletitle{The Unreasonable Effectiveness of Deep Features as
  a Perceptual Metric}. In \bibinfo{booktitle}{\emph{2018 IEEE/CVF Conference
  on Computer Vision and Pattern Recognition}}. \bibinfo{pages}{586--595}.
\newblock
\href{https://doi.org/10.1109/CVPR.2018.00068}{doi:\nolinkurl{10.1109/CVPR.2018.00068}}


\end{thebibliography}
